\documentclass[aps,prb,fleqn,12pt,superscriptaddress]{revtex4}
\usepackage{epsfig,color}
\usepackage{graphicx}
\usepackage{pdfpages}
\usepackage{braket}
\usepackage{amssymb,amsmath}
\usepackage{hyperref}
\begin{document}
\title{Spin-orbit coupling and magnetism in {\boldmath $\rm Sr_2CrO_4$}} 
\author{Shubhajyoti Mohapatra}
\affiliation{Department of Physics, Indian Institute of Technology, Kanpur - 208016, India}
\author{Dheeraj Kumar Singh}
\affiliation{Thapar Institute of Engineering and Technology, Patiala - 147004, India}
\author{Avinash Singh}
\email{avinas@iitk.ac.in}
\affiliation{Department of Physics, Indian Institute of Technology, Kanpur - 208016, India}
\date{\today} 
\begin{abstract}
With octahedrally coordinated $t_{\rm 2g}$ orbitals which are active at filling $n=2$, the $\rm Sr_2CrO_4$ compound exhibits rich interplay of spin-orbital physics with tetragonal distortion induced crystal field tuning by external agent such as pressure. Considering both reversed and restored crystal field regimes, collective spin-orbital excitations are investigated in the antiferromagnetic state  using the generalized self consistent + fluctuations approach including spin-orbit coupling (SOC). A transition is found from staggered to entangled orbital order at critical SOC value in the realistic regime. Behavior of the calculated energy scales of collective excitations with crystal field is in striking similarity to that of the transition temperatures with pressure as obtained from susceptibility and resistivity anomalies in high-pressure studies. 
\end{abstract}
\maketitle
\newpage

\section{Introduction}
Among layered perovskite structured $3d$ transition metal compounds $\rm Sr_2MO_4$ where M=V,Cr,Mn with $n=1,2,3$ electrons in the $t_{\rm 2g}$ sector, $\rm Sr_2CrO_4$ presents a promising case for investigating spin-orbit coupling (SOC) and Coulomb interaction effects in view of the rich spin-orbital physics exhibited by this compound. Among other members of the Ruddlesden-Popper (RP) series $\rm Sr_{n+1}Cr_n O_{3n+1}$, the cubic perovskite $\rm SrCrO_3$ ($n=\infty$) is known to exhibit weak antiferromagnetic (AFM) order accompanied with small tetragonal distortion and staggered $yz/xz$ orbital order due to the $(xy)^1(yz,xz)^1$ electronic configuration.\cite{zhou_PRL_2006, martin, lee1_PRB_2009} The AFM state in the bilayer compound $\rm Sr_3Cr_2O_7$ ($n = 2$) is also of a similar nature.\cite{jeanneau_PRL_2017, aligia_PRB_2019} With $n=2$ also in the cubic perovskite $\rm LaVO_3$ exhibiting active orbital degree of freedom, inter-site spin-orbital entanglement was found to play an important role in magnetic and orbital ordering temperature, optical spectra, and magnon spectra anomalies.\cite{oles_JPCM_2012} 

It is only recently that pure samples of $\alpha$-Sr$_2$CrO$_4$ have been synthesized successfully in bulk, and uncovering the interesting magnetic and orbital physics has attracted considerable interest. Magnetic susceptibility and specific heat measurements, as well as neutron scattering and muon spin rotation studies, have indicated phase transitions at temperatures $\sim$ 110 K and 140 K.\cite{sakurai_JPSJ_2014,nozaki_JPS_2018,jeanneau_EPL_2019,zhu_arxiv_2021} A high-pressure study of $\rm Sr_2CrO_4$ has been carried out recently using magnetic susceptibility and electrical resistivity measurements up to 14 GPa pressure.\cite{yamauchi_PRL_2019} The evolution of the 110 K and 140 K transitions was investigated with pressure in the $P \lesssim 3$ GPa regime where the transition temperatures were identified based on $\chi-T$ anomalies. Similar successive transitions in temperature based on $\rho-T$ anomalies were also identified in the $P \gtrsim 6$ GPa regime where the crystal field is restored.  

With a $\rm K_2NiF_4$-type crystal structure, the $\rm CrO_6$ octahedra in $\alpha$-Sr$_2$CrO$_4$ are elongated along $c$ axis.\cite{sakurai_JPSJ_2014,baikie_JSSC_2007} The octahedral crystal field partially lifts the five-fold degenerancy of $3d$ orbitals into $e_g$ and $t_{2g}$ states. Due to octahedral elongation, the low-lying $t_{2g}$ states are conventionally expected to further split into low-lying degenerate $yz,xz$ levels and higher-energy $xy$ level.\cite{weng_PRB_2006} However, the $xy$ orbital was found to be low-lying instead in the more recent DFT study.\cite{ishikawa_JPSJ_2017} This counter-intuitive reversal of usual crystal field effect was attributed to negative charge-transfer gap associated with Cr-3$d$ and O-2$p$ orbitals. Because of the competing ionic and covalent bonding interactions, the $t_{2g}$ set was projected to become degenerate only for highly elongated $\rm CrO_6$ octahedra, whereas for realistic elongation in $\alpha$-Sr$_2$CrO$_4$, the $xy$ orbital was found to be relatively lower in energy.\cite{jeanneau_EPL_2019} 

Theoretically, spin-orbital physics in Sr$_2$CrO$_4$ is yet to be fully understood as only few studies have been carried out. These include: density functional theory (DFT) based calculations,\cite{weng_PRB_2006,ishikawa_JPSJ_2017,takahashi_JPSC_2020,pandey_PRB_2021} tight-binding model fitting to DFT band structure and real-space Hartree-Fock (HF) approximation, supplemented by density-matrix renormalization group (DMRG) study.\cite{pandey_PRB_2021} The Cr$^{4+}$ ion has $3d^2$ configuration, and with one electron in the low-lying $xy$ orbital, the system is rendered orbitally active due to one remaining electron in the degenerate $yz,xz$ sector. Staggered ($\pi,\pi$) magnetic and orbital order was obtained in the HF study for large on-site Coulomb interaction. However, the study was carried out for small system size and the full range of crystal field splitting was not considered. The low-energy collective spin-orbital excitations remain largely unexplored both experimentally as well as theoretically. Understanding the observed behavior of transition temperatures in high pressure studies of $\rm Sr_2CrO_4$, where conventional crystal field is restored, also remains elusive.

Several important physical effects have not been considered in earlier theoretical works. These include: SOC induced magnetic anisotropy, orbital moments, Coulomb interaction induced orbital mixing terms, and SOC renormalization. In view of the crystal field tuning by pressure, and the peculiar insulator-insulator transition in the crossover region between reversed and restored crystal field regimes,\cite{yamauchi_PRL_2019,takahashi_JPSC_2020} investigation of the orbital mixing effect is of crucial importance to understand how the insulating gap is protected even as the $xy$ level crosses the $yz,xz$ levels.

Due to the rich interplay expected between SOC, crystal field, and Coulomb interaction terms, investigation of composite spin-orbital ordering, SOC induced magnetic anisotropy, and collective spin-orbital excitations is of particular interest for the $\rm Sr_2CrO_4$ compound. For this purpose, we will employ the generalized self consistent + fluctuations approach in which all physical elements are treated on equal footing. This approach was recently used to investigate spin-orbital fluctuations in several $4d$ and $5d$ compounds ($\rm NaOsO_3$, $\rm Ca_2RuO_4$, $\rm Sr_2IrO_4$) with $n=3,4,5$ electrons in the $\rm t_{2g}$ sector.\cite{mohapatra_JPCM_2020,mohapatra_JPCM_2021} 

The structure of this paper is as below. The realistic model Hamiltonian and the generalized self-consistent + fluctuations approach are briefly reviewed in Sections II and III. Results are presented in Sections IV and V for the staggered and entangled orbital orders. Important features of our calculated results, including comparison of the behavior of excitation energy scales with crystal field and that of the measured transition temperatures with pressure, are discussed in Sec. VI. Finally, conclusions are presented in Sec. VII. 

\section{Model Hamiltonian}
We consider the Hamiltonian ${\cal H} = {\cal H}_{\rm cf} + {\cal H}_{\rm band} + {\cal H}_{\rm SOC} + {\cal H}_{\rm int}$ within the $t_{\rm 2g}$ manifold consisting of the three-orbital ($\mu=yz,xz,xy$) and two-spin ($\sigma=\uparrow,\downarrow$) basis defined with respect to common spin-orbital coordinate axes along planar Cr-O-Cr directions. The crystal field and band terms have been discussed earlier,\cite{mohapatra_JPCM_2020} and are briefly summarized below.

The $xy$ orbital energy offset $\epsilon_{xy}$ (relative to degenerate $yz/xz$ orbitals) represents the effective crystal field splitting. The first, second, and third neighbor hopping terms for $xy$ orbital are represented by $t_1$, $t_2$, $t_3$, respectively. For $xz$ orbital, $t_4$ and $t_5$ are the nearest-neighbor (NN) hopping terms in $x$ and $y$ directions, and similarly for the $yz$ orbital. We have taken hopping parameter values: ($t_1$, $t_2$, $t_3$, $t_4$, $t_5$)=$(-1.0, 0.3, 0, -1.0, 0.2)$, and considered the range $-1\leq \epsilon_{xy} \leq +1$, all in units of the realistic hopping energy scale $|t_1|$=250 meV.\cite{pandey_PRB_2021} As there is no experimental evidence for octahedral rotation/tilting in $\rm Sr_2CrO_4$, the orbital mixing hopping terms have been set to zero.  

For the bare spin-orbit coupling term, we consider the representation in spin space:
\begin{eqnarray} 
{\cal H}_{\rm SOC} (i) & = & -\lambda {\bf L} \cdot {\bf S} = -\lambda (L_z S_z + L_x S_x + L_y S_y) \nonumber \\ 
&=& \sum_\alpha \begin{pmatrix} \psi_{\mu \uparrow}^\dagger & \psi_{\mu \downarrow}^\dagger \end{pmatrix} \begin{pmatrix} i \sigma_\alpha \lambda /2 \end{pmatrix} 
\begin{pmatrix} \psi_{\nu \uparrow} \\ \psi_{\nu \downarrow} 
\end{pmatrix} + {\rm H.c.}
\label{soc}
\end{eqnarray}
where the orbital pair $(\mu,\nu)=(yz,xz)$ for spin component $\alpha=z$, and similarly for other components. Since spin rotation symmetry is explicitly broken, the SOC term therefore generates anisotropic magnetic interactions from its interplay with other Hamiltonian terms. 

Finally, for the Coulomb interaction part of the Hamiltonian, we consider (for site $i$):
\begin{equation}
{\cal H}_{\rm int} (i) = U\sum_{\mu}{n_{i\mu\uparrow}n_{i\mu\downarrow}} + U^{\prime \prime}\sum_{\mu<\nu} n_{i\mu} n_{i\nu} - 2J_{\mathrm H} \sum_{\mu<\nu} {\bf S}_{i\mu}.{\bf S}_{i\nu} +J_{\mathrm P} \sum_{\mu \ne \nu} a_{i \mu \uparrow}^{\dagger} a_{i \mu\downarrow}^{\dagger}a_{i \nu \downarrow} a_{i \nu \uparrow} 
\label{h_int}
\end{equation} 
including the intra-orbital $(U)$ and inter-orbital $(U^{\prime\prime}=U-5J_{\rm H}/2)$ density interaction terms, along with Hund's coupling $(J_{\rm H})$, and pair hopping $(J_{\rm P}=J_{\rm H})$ interaction terms. Here $a_{i\mu\sigma}^{\dagger}$ and $a_{i\mu \sigma}$ are the electron creation and annihilation operators for site $i$, orbital $\mu$, spin $\sigma=\uparrow ,\downarrow$. The density operator $n_{i\mu\sigma}=a_{i\mu\sigma}^\dagger a_{i\mu\sigma}$, total density operator $n_{i\mu}=n_{i\mu\uparrow}+n_{i\mu\downarrow}=\psi_{i\mu}^\dagger \psi_{i\mu}$, and spin density operator ${\bf S}_{i\mu} = \psi_{i\mu}^\dagger ${\boldmath $\sigma$}$ \psi_{i\mu}$ in terms of the electron field operator $\psi_{i\mu}^\dagger=(a_{i\mu\uparrow}^{\dagger} \; a_{i\mu\downarrow}^{\dagger})$. All interaction terms above are SU(2) invariant and thus possess spin rotation symmetry.  

In the following, we will consider $U$ in the range 8-12 (2-3 eV), $J_{\rm H}/U$ between 1/10 to 1/6, and bare SOC values $\lambda \leq  0.2$, all in the energy scale unit $|t_1|$=250 meV. The limiting SOC value (50 meV) lies in the realistic range for $3d$ elements.\cite{radwanski_APP_2000,stohr_2006} 

\section{Generalized self consistent + fluctuations approach}

In the generalized self consistent approach, besides the standard orbital diagonal terms, all terms involving orbital mixing (OM) condensates of the spin and charge operators are also included, resulting in additional orbital mixing contributions: 
\begin{equation}
[{\cal H}_{\rm int}^{\rm HF}]_{\rm OM} = \sum_{i,\mu < \nu} \psi_{i\mu}^{\dagger} \left [
-\makebox{\boldmath $\sigma . \Delta$}_{i\mu\nu} + {\cal E}_{i\mu\nu} {\bf 1} \right ] \psi_{i\nu} + {\rm H.c.}
\label{h_hf_od} 
\end{equation}  
of the Coulomb interaction terms (Eq. \ref{h_int}) in the Hartree-Fock (HF) approximation. The orbital mixing spin and charge fields are self-consistently determined from:
\begin{eqnarray}
2\makebox{\boldmath $\Delta$}_{i\mu\nu} &=& (U'' + J_{\rm H}/2) \langle \makebox{\boldmath $\sigma$}_{i\nu\mu} \rangle + J_{\rm P} \langle \makebox{\boldmath $\sigma$}_{i\mu\nu} \rangle \nonumber \\
2{\cal E}_{i\mu\nu} &=& (-U'' + 3J_{\rm H}/2) \langle n_{i\nu\mu} \rangle + J_{\rm P} \langle n_{i\mu\nu} \rangle  
\label{sc_od}
\end{eqnarray}
in terms of the corresponding condensates $\langle \makebox{\boldmath $\sigma$}_{i\mu\nu}\rangle \equiv \langle \psi_{i\mu}^{\dagger} \makebox{\boldmath $\sigma$} \psi_{i\nu} \rangle$ and $\langle n_{i\mu\nu} \rangle \equiv \langle \psi_{i\mu}^{\dagger} {\bf 1} \psi_{i\nu} \rangle$. The orbital mixing terms above explicitly preserve spin rotation symmetry, and are generally finite due to orbital mixing induced by SOC or octahedral tilting/rotation. In the following, we will see that these terms can also be generated spontaneously. 

The orbital mixing charge and spin condensates yield the orbital moments and Coulomb interaction induced SOC renormalization: 
\begin{eqnarray}
\langle L_\alpha \rangle & = & -i \left [ \langle \psi_\mu^\dagger \psi_\nu\rangle - \langle \psi_\mu^\dagger \psi_\nu\rangle^* \right ] = 2\ {\rm Im}\langle \psi_\mu^\dagger \psi_\nu\rangle \nonumber \\
\lambda^{\rm int}_\alpha & = & (U'' - J_{\rm H}/2) {\rm Im}\langle \psi_\mu^\dagger \sigma_\alpha \psi_\nu\rangle
\label{phys_quan}
\end{eqnarray}
where the orbital pair ($\mu,\nu$) corresponds to the component $\alpha=x,y,z$. The last equation yields the Coulomb renormalized SOC values $\lambda_\alpha = \lambda + \lambda_\alpha^{\rm int}$ where $\lambda$ is the bare SOC value. 

The above approach was recently applied to several $4d$ and $5d$ compounds ($\rm NaOsO_3$, $\rm Ca_2RuO_4$, $\rm Sr_2IrO_4$) with electron fillings $n=3,4,5$ in the $t_{\rm 2g}$ sector.\cite{mohapatra_JPCM_2020,mohapatra_JPCM_2021} Recent application to the vanadate compound $\rm Sr_2VO_4$ with $n=1$ reveals a SOC induced transition from staggered orbital + FM order to entangled orbital + AFM order.\cite{vanadate_2021} Generally, coupling of orbital moments to orbital fields and interaction-induced SOC renormalization effects highlight the important role of orbital mixing condensates in the emergent spin-orbital physics.\cite{mohapatra_JMMM_2021} 

For investigation of collective spin-orbital excitations, we consider the time-ordered generalized fluctuation propagator:
\begin{equation}
[\chi({\bf q},\omega)] = \int dt \sum_i e^{i\omega(t-t')} 
e^{-i{\bf q}.({\bf r}_i -{\bf r}_j)} 
\times \langle \Psi_0 | T [\sigma_{\mu\nu}^\alpha (i,t) \sigma_{\mu'\nu'}^{\alpha'} (j,t')] |\Psi_0 \rangle 
\end{equation}
in the self-consistent ground state $|\Psi_0 \rangle$. The generalized spin-charge operators at lattice sites $i,j$ are defined as $\sigma_{\mu\nu}^\alpha = \psi_\mu ^\dagger \sigma^\alpha \psi_\nu$, with $\sigma^\alpha$ defined as Pauli matrices for $\alpha=x,y,z$ and unit matrix for $\alpha=c$. Including these operators ensures consistency with the self consistent approach where all generalized spin $\langle \psi_\mu ^\dagger \makebox{\boldmath $\sigma$}\psi_\nu \rangle$ and charge $\langle \psi_\mu ^\dagger \psi_\nu \rangle$ condensates were included.

The generalized fluctuation propagator in the random phase approximation (RPA) was investigated recently for several $4d$ and $5d$ compounds with electron fillings $n=3,4,5$ in the $t_{\rm 2g}$ sector.\cite{mohapatra_JPCM_2021} Since the generalized spin and charge operators $\psi_\mu ^\dagger \sigma^\alpha \psi_\nu$ include spin ($\mu=\nu$, $\alpha=x,y,z$), orbital ($\mu\ne\nu$, $\alpha=c$), and spin-orbital ($\mu\ne\nu$, $\alpha=x,y,z$) cases, the spectral function of the fluctuation propagator:
\begin{equation}
{\rm A}_{\bf q}(\omega) = \frac{1}{\pi} {\rm Im \; Tr} [\chi({\bf q}, \omega)]_{\rm RPA}
\label{spectral}
\end{equation}
provides information about the various collective excitations (magnon, orbiton, and spin-orbiton). Orbiton and spin-orbiton modes correspond to same-spin and spin-flip particle-hole excitations involving different orbitals. 

Due to the active $yz/xz$ orbital degree of freedom in $\rm Sr_2CrO_4$ in the $\epsilon_{xy}\sim -1$ regime (where $n_{xy} \approx 1$ and $n_{yz}+n_{xz} \approx 1$), we will show that there exists a critical SOC value $\lambda^*$ (weakly $U$ and J$_{\rm H}$ dependent) which determines the nature of the orbital ordering. The AFM order (dominantly due to $xy$ orbital electron) is accompanied with either staggered orbital order (for $\lambda < \lambda^*$) or with entangled orbital order (for $\lambda > \lambda^*$). Results of the self consistent calculation for these two cases are discussed in the next two sections. 

\newpage
\begin{figure}
\vspace*{0mm}
\hspace*{0mm}
\psfig{figure=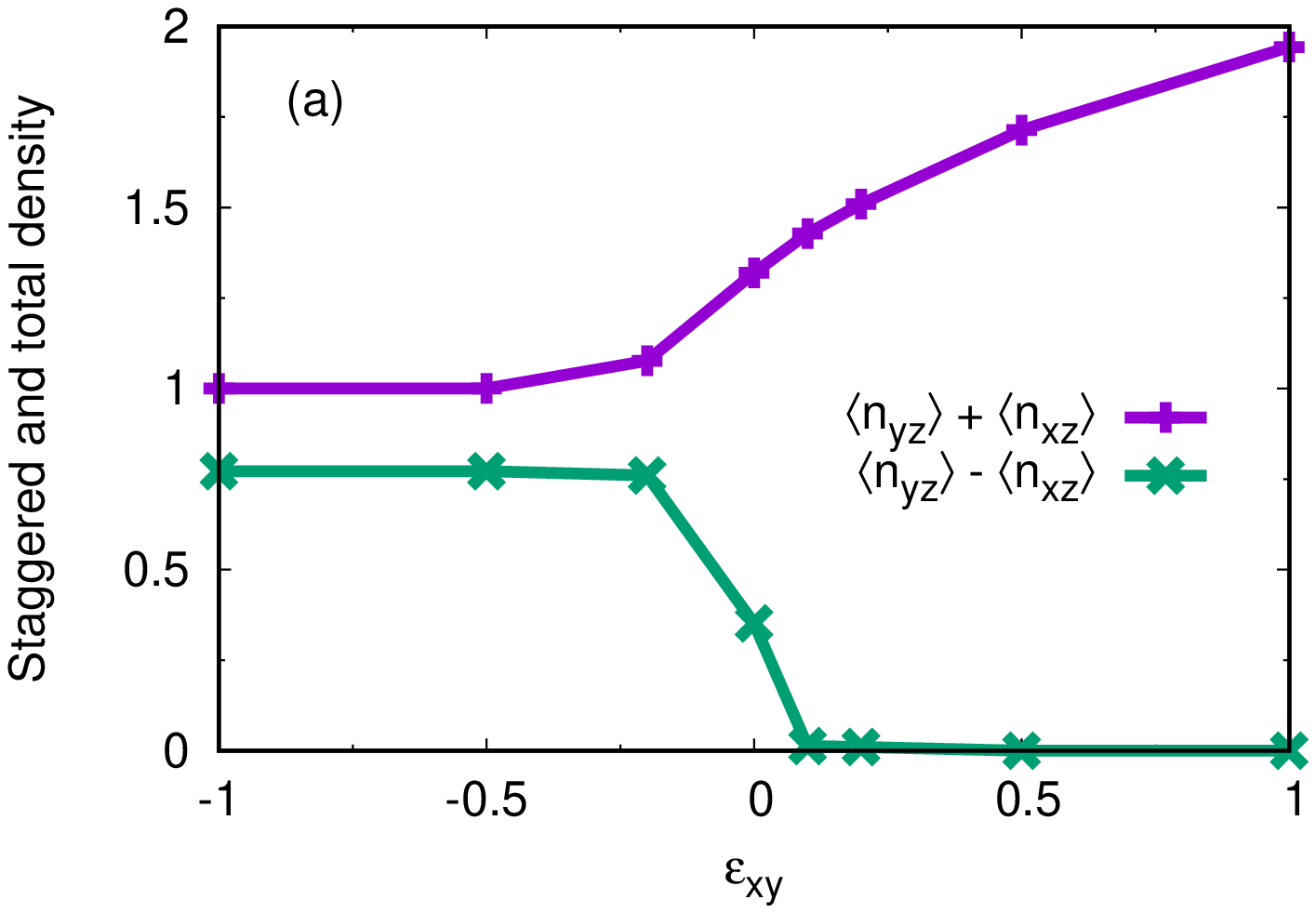,angle=-90,width=65mm}
\psfig{figure=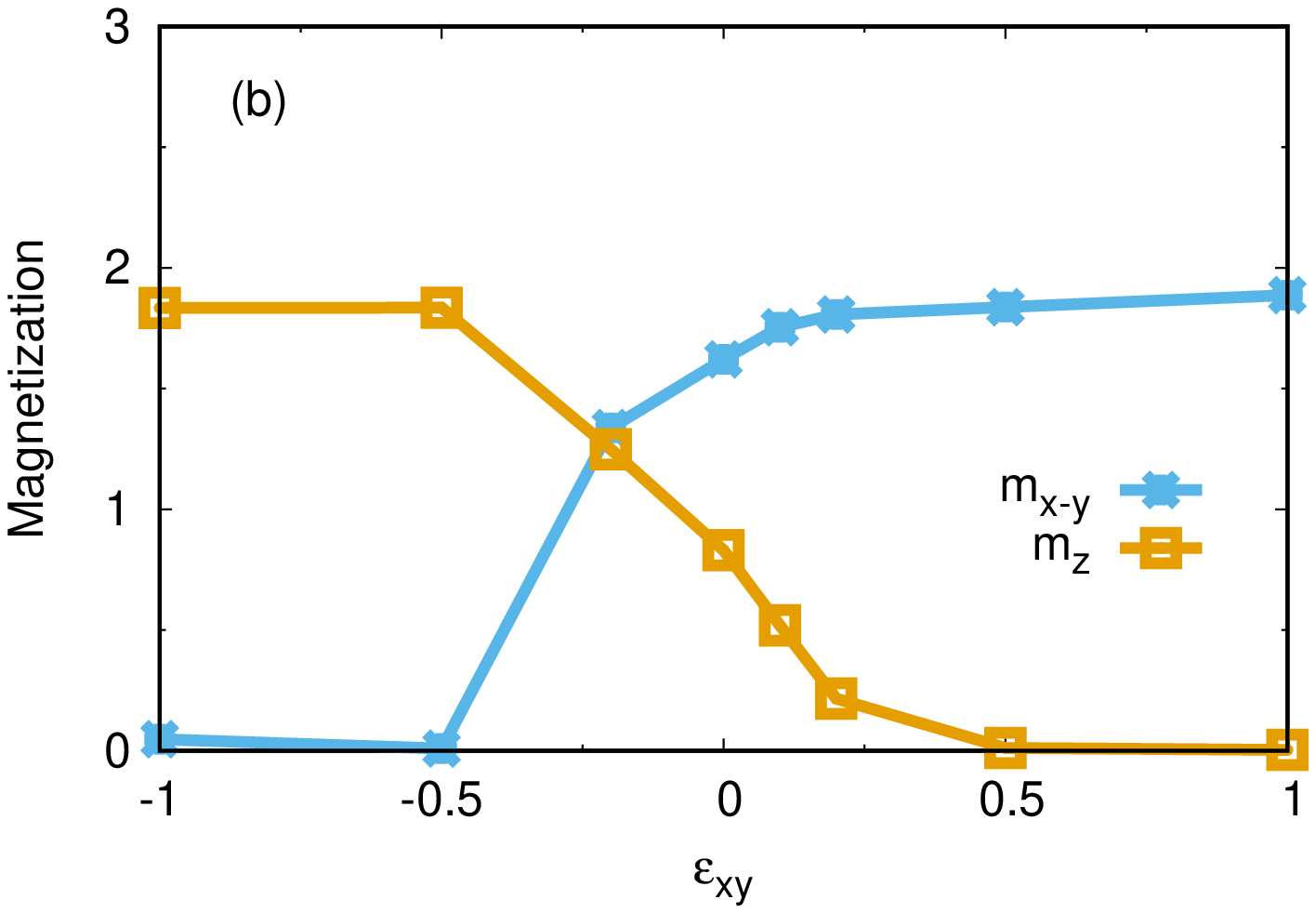,angle=-90,width=65mm}
\psfig{figure=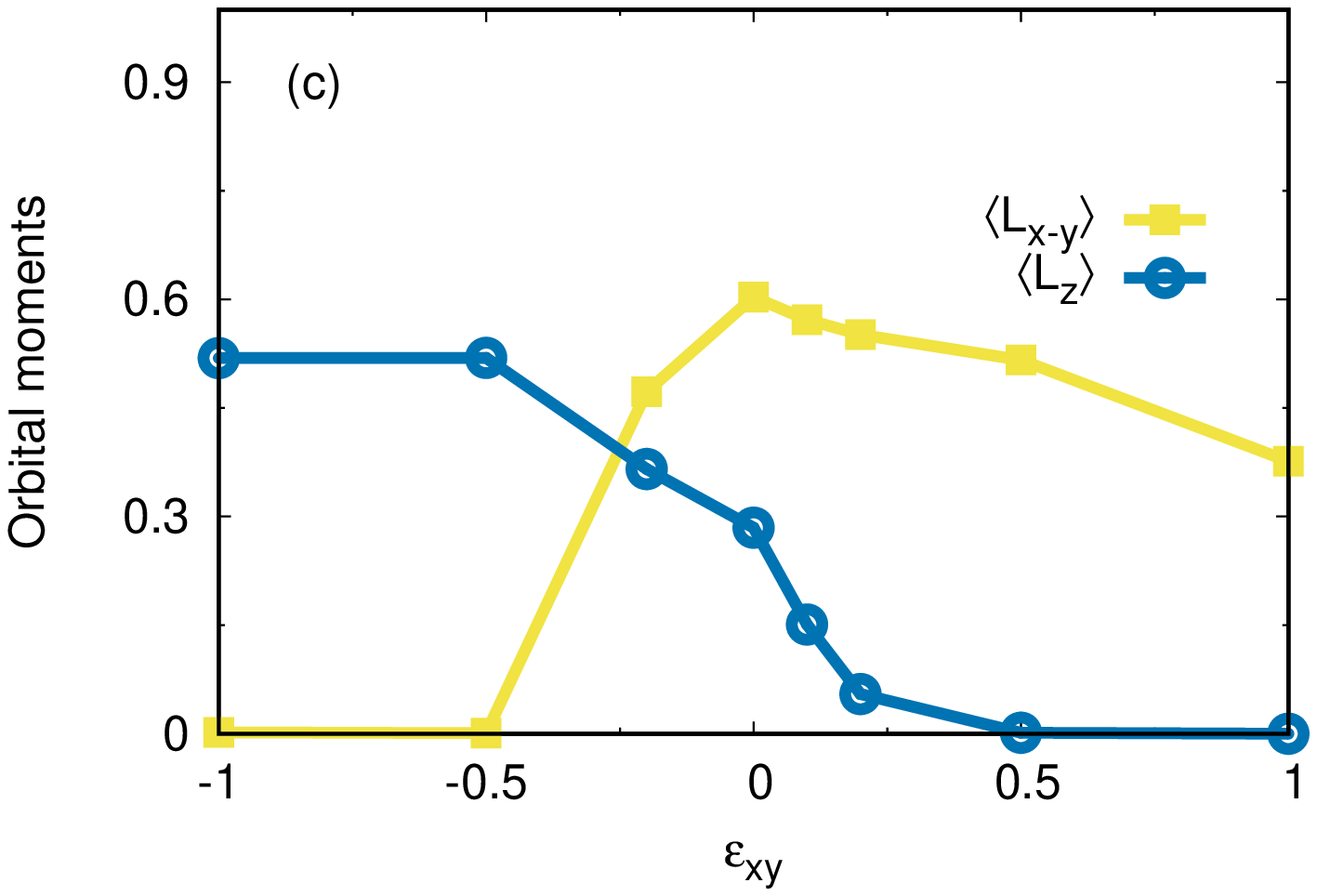,angle=-90,width=65mm}
\psfig{figure=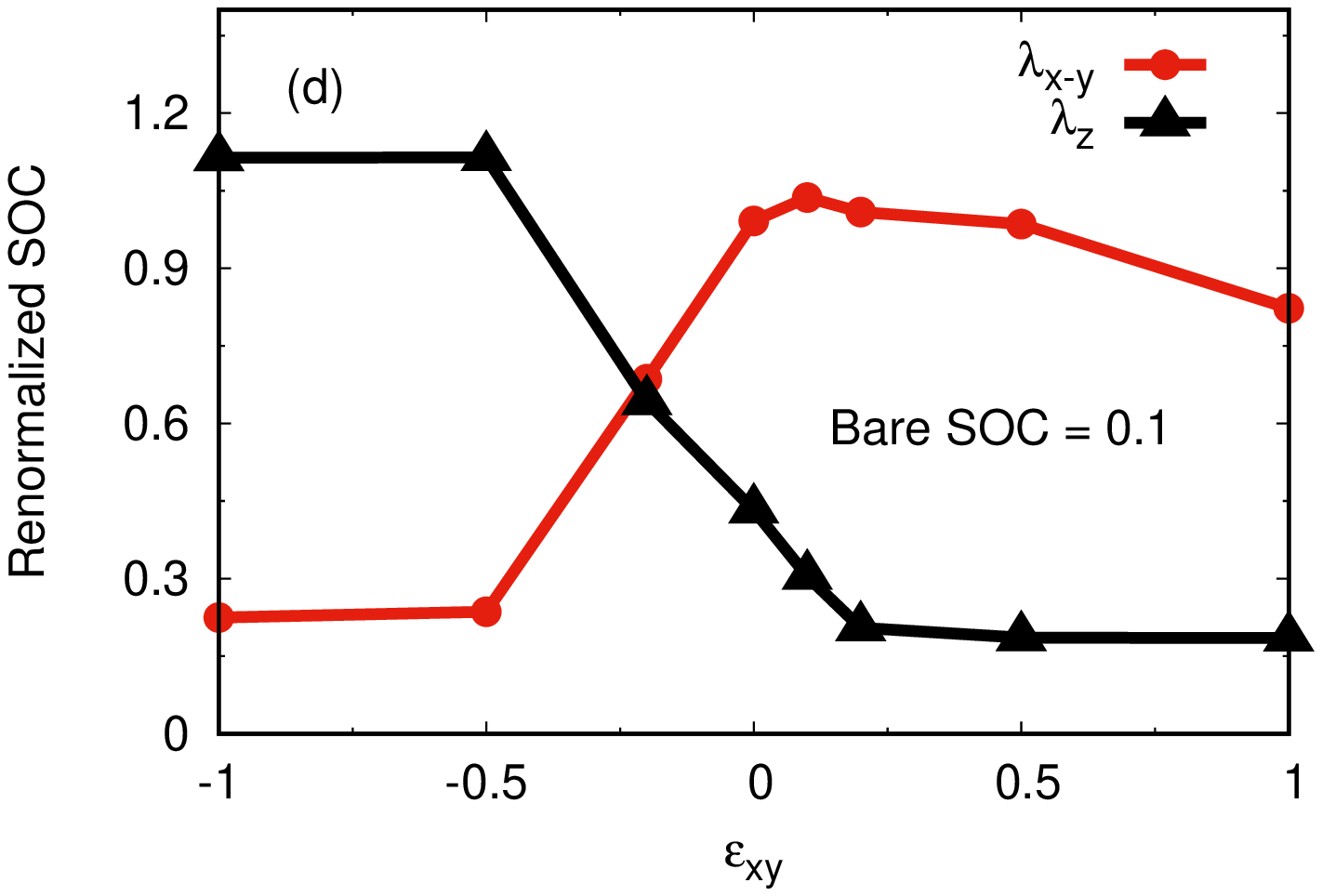,angle=-90,width=65mm}
\caption{(a) The total and differential electron occupancies in the $yz/xz$ sector, showing strong proclivity for staggered $(\pi,\pi)$ structure when $n_{yz}+n_{xz} \sim 1$. Axial and planar magnetic moments (b) and orbital moments (c), showing the magnetic reorientation transition. Ordering direction switches from $z$ to $x-y$ plane near $\epsilon_{xy}=-0.25$. (d) Coulomb renormalized SOC values show similar behavior.} 
\label{fig1}
\end{figure}

\section{Staggered Orbital Order ($\lambda < \lambda^*$)}
\subsection{Magnetic reorientation transition}
\begin{figure}
\vspace*{0mm}
\hspace*{0mm}
\psfig{figure=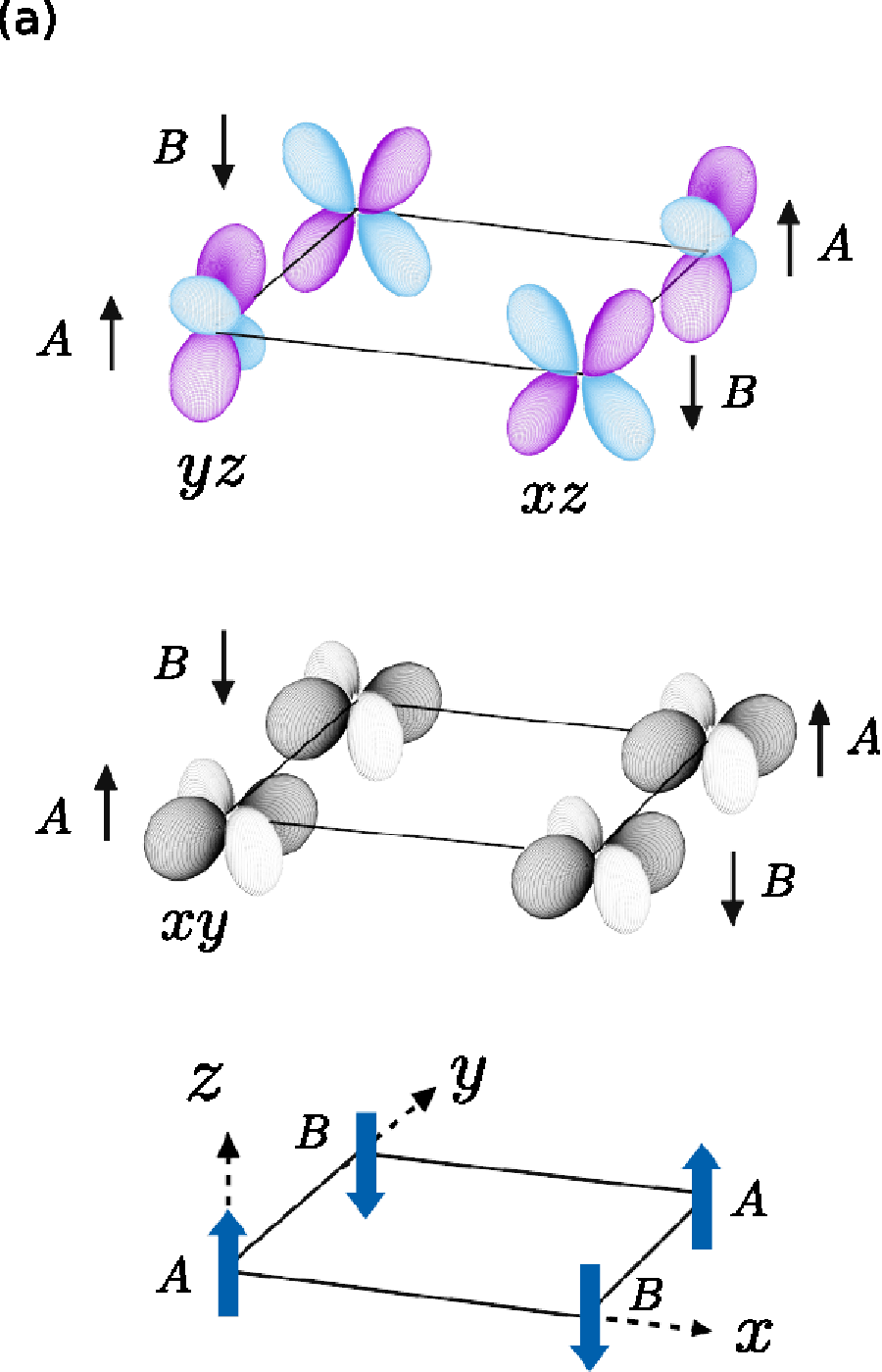,angle=0,width=40mm}\hspace*{25mm}
\psfig{figure=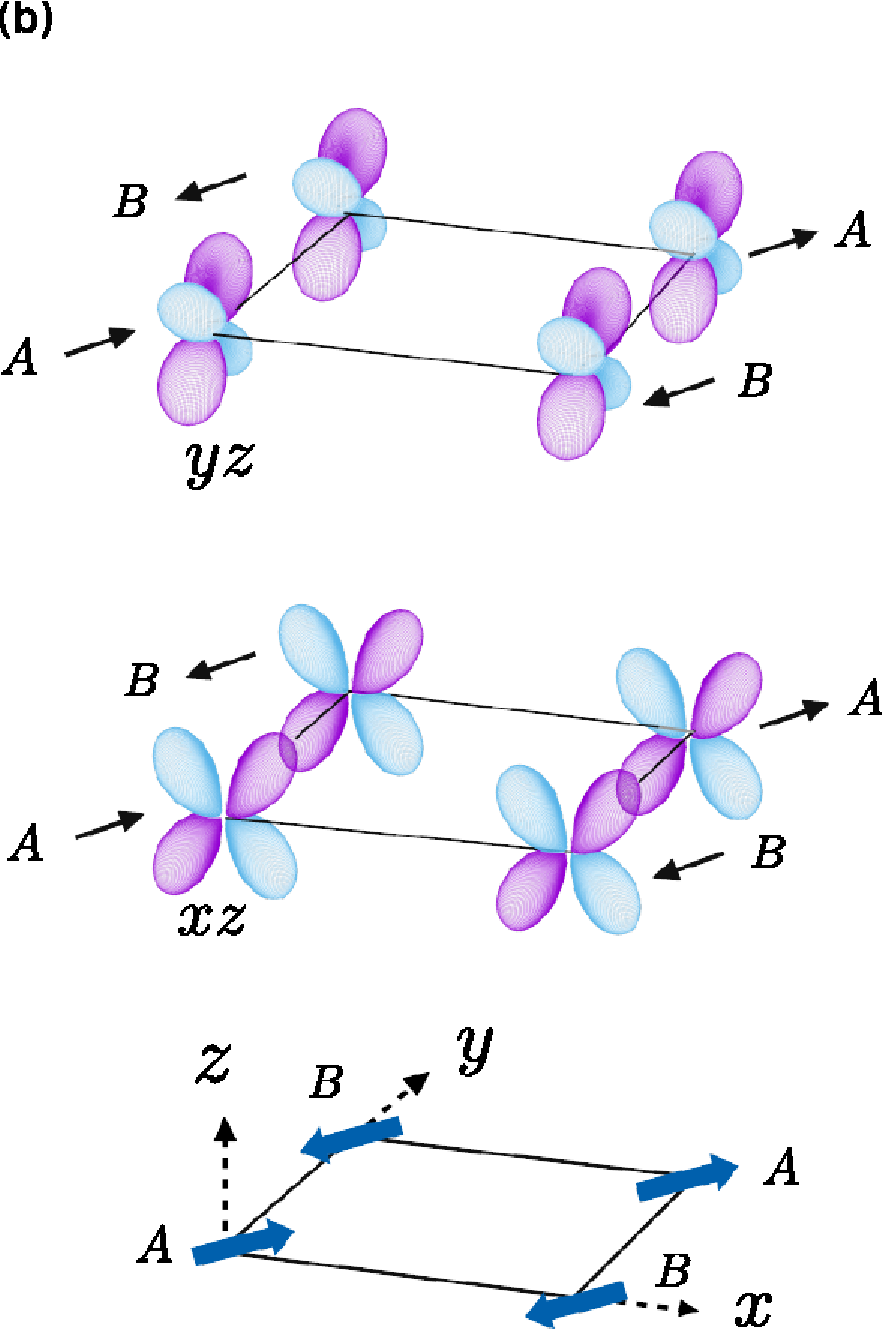,angle=0,width=40mm}
\caption{Schematic representation of the composite spin-orbital structure showing AFM order (a) along $z$ direction with staggered $yz/xz$ ordering (for $\epsilon_{xy}\sim -1$), and (b) along azimuthal angle $\phi=\pi/4$ in the $x-y$ plane when both $yz,xz$ orbitals are occupied (for $\epsilon_{xy}\sim +1)$.} 
\label{fig2}
\end{figure}

Throughout this section we will consider parameter values: $U=8$, $J_{\rm H}=U/6$, and $\lambda=0.1$, unless otherwise mentioned. Results of the self consistent calculation are shown in Fig. \ref{fig1}. Besides the spin magnetic moments, the orbital magnetic moments and Coulomb renormalized SOC values are also shown for different $\epsilon_{xy}$ values. The staggered magnetization values shown in Fig. \ref{fig1}(b) refer to orbital-summed $z$ and $x-y$ plane components:
\begin{eqnarray}
m_z &=& \sum_\mu m_\mu ^z = \sum_\mu (n_\mu ^\uparrow - n_\mu ^\downarrow)_A = \sum_\mu (n_\mu ^\downarrow - n_\mu ^\uparrow)_B \nonumber \\
m_{x-y} &=& \sum_\mu [(m_\mu ^x)^2 + (m_\mu ^y)^2]^{1/2}_{A/B}
\end{eqnarray}
in the AFM order. The planar orbital moment and renormalized SOC components $\langle L_{x-y} \rangle$ and $\lambda_{x-y}$ were also evaluated as above (without the orbital sum). The magnetic and orbital moments are seen to undergo a reorientation transition. The composite spin-orbital structure for the two limiting cases $\epsilon_{xy} \sim \pm 1$ is illustrated in Fig. \ref{fig2}, and the salient features are discussed below.

\subsection{Staggered orbital + AFM ($z$) order}
For $\epsilon_{xy}\sim -1$ (reversed crystal field), the $xy$ band is pulled down, resulting in nominally $(xy)^1(yz,xz)^1$ electron occupancy. The $yz,xz$ sector develops staggered $(\pi,\pi)$ orbital order, which is stabilized by the density interaction term $U''$. Equivalence between attributing this structure either to density interaction or to Jahn-Teller effect has been reviewed earlier in the context of manganites with two $e_{\rm g}$ orbitals and one electron per site.\cite{dheeraj_JPCM_2010} 

The AFM order is stabilized by interactions between $xy$ moments resulting from the $xy$ orbital electron hopping; the $yz,xz$ moments order accordingly due to Hund's coupling. The AFM order is frustrated by the preferred FM order of $yz,xz$ moments.\cite{roth_PR_1966} The energy gain due to virtual electron hopping [$\sim t_4^2/(U''-J_{\rm H})$] is greater for parallel neighboring spins since the doubly occupied intermediate state has lower energy due to Hund's coupling. The relative energy gain for parallel vs. antiparallel spins ($\sim t_4^2 J_{\rm H}/U^2$) becomes important when $U$ is not large. Due to SOC induced magnetic anisotropy, the magnetic and orbital moments are oriented along the $z$ direction. 

\subsection{AFM ($x-y$) order}
For $\epsilon_{xy}\sim +1$ (conventional crystal field), the $xy$ band is pushed up, aided by the Coulomb renormalization of crystal field splitting, resulting in nominally $(xy)^0(yz,xz)^2$ electron occupancy. The AFM order is dominantly stabilized by AFM interaction between $xz$ ($yz$) moments in $x$ $(y)$ direction. Due to the SOC induced easy-plane magnetic anisotropy, the $yz,xz$ magnetic moments as well as the orbital moments order in the $x-y$ plane. Both orders are oriented along azimuthal angle $\phi=\pi/4$ within the $x-y$ plane.

\subsection{SOC induced anisotropic interactions} 
The SOC induced anisotropic magnetic interactions provide insight into how the onset of staggered $yz/xz$ ordering (for $\epsilon_{xy}\sim -1$) is responsible for reorienting the AFM order along the $z$ direction from the easy-plane orientation when both $yz,xz$ orbitals are occupied (for $\epsilon_{xy}\sim +1$). In analogy with the strong-coupling analysis for the spin-dependent hopping terms $i${\boldmath $\sigma . t'_{ij}$},\cite{hc_JMMM_2019} strong-coupling expansion for the bare SOC terms $-\lambda\sum_{\alpha=x,y,z} L_\alpha S_\alpha$ to second order in $\lambda$ yields the anisotropic diagonal (AD) intra-site interactions (for site $i$): 
\begin{eqnarray}
[H^{(2)}_{\rm eff}]_{\rm AD} (i) &=& \frac{4 (\lambda/2)^2 }{U} \left [ S_{yz}^z S_{xz}^z - (S_{yz}^x S_{xz}^x  + S_{yz}^y S_{xz}^y) \right ] \nonumber \\
&+& \frac{4 (\lambda/2)^2 }{U} \left [S_{xz}^x S_{xy}^x - (S_{xz}^y S_{xy}^y  + S_{xz}^z S_{xy}^z) \right ] \nonumber \\
&+& \frac{4 (\lambda/2)^2 }{U} \left [S_{xy}^y S_{yz}^y - (S_{xy}^x S_{yz}^x  + S_{xy}^z S_{yz}^z) \right ] 
\label{h_eff}
\end{eqnarray}
between the $yz,xz,xy$ moments. Here all three orbitals have been assumed to be nominally half-filled in this general analysis. 

Now, for $\epsilon_{xy}\sim +1$, when only $yz,xz$ moments are present, only the first term in Eq. (\ref{h_eff}) is operative, which directly yields preferential $x-y$ plane ordering (easy-plane anisotropy) for parallel $yz,xz$ moments (enforced by Hund's coupling). On the other hand, for $\epsilon_{xy}\sim -1$, when only $xy$ and $yz$ moments are present on A sublattice (say) and only $xy$ and $xz$ moments on B sublattice, only the third and second terms are operative, respectively. There is no frustration if the moments are oriented along $z$ direction, whereas if moments are oriented along $x$ direction on A sublattice and along $y$ direction on B sublattice, the AFM order is frustrated. The resulting easy ($z$) axis anisotropy thus involves a crucial interplay between the SOC induced anisotropic interactions, orbital occupancy, Hund's coupling, and AFM Heisenberg interactions. 

In the planar ordering case ($\epsilon_{xy}\sim +1$), the continuous symmetry for in-plane spin rotation is further reduced to $C_4$ symmetry due to weak easy-axis anisotropy (along the $45^\circ$ directions) resulting from weak anisotropic interactions (second and third terms in Eq. \ref{h_eff}) due to the small $xy$ moment in this regime.\cite{mohapatra_JPCM_2021} The above qualitative analysis for the SOC induced magnetic anisotropy and ordering directions in the two opposite $\epsilon_{xy}$ regimes is confirmed by the generalized self consistent calculations described earlier. 

\begin{figure}
\vspace*{0mm}
\hspace*{0mm}
\psfig{figure=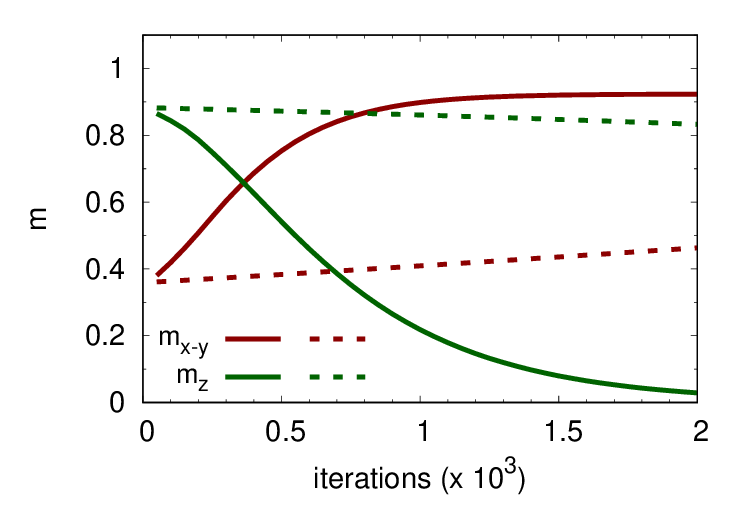,angle=0,width=80 mm,angle=0}
\caption{Variation of the axial ($m_z$) and planar ($m_{x-y}$) sublattice magnetization components with iterations in the (i) standard (dashed line) and (ii) generalized (solid line) self consistent calculations. Starting with nearly-$z$ initial direction, the AFM ordering direction approaches the $x-y$ plane. Here bare SOC value $\lambda=0.1$, $U=8$, $\epsilon_{xy} = +1$.} 
\label{fig3}
\end{figure}

As an illustration of the difference between standard and generalized self consistent approaches, Fig. \ref{fig3} shows the evolution of axial and planar sublattice magnetization components (identical for both $yz,xz$ orbitals) with iterations in the easy-plane anisotropy case ($\epsilon_{xy}\sim +1$), starting with nearly-$z$ initial ordering direction. The extremely low magnetic anisotropy energy scale $\lambda^2/U$ in Eq. (\ref{h_eff}) is reflected in the extremely slow approach of the AFM ordering direction to the $x-y$ plane. The approach is relatively much faster in the generalized (compared to the standard) self consistent calculation, showing effectively enhanced Coulomb renormalized SOC effect on magnetic anisotropy. 


\begin{figure}
\vspace*{-10mm}
\hspace*{0mm}
\psfig{figure=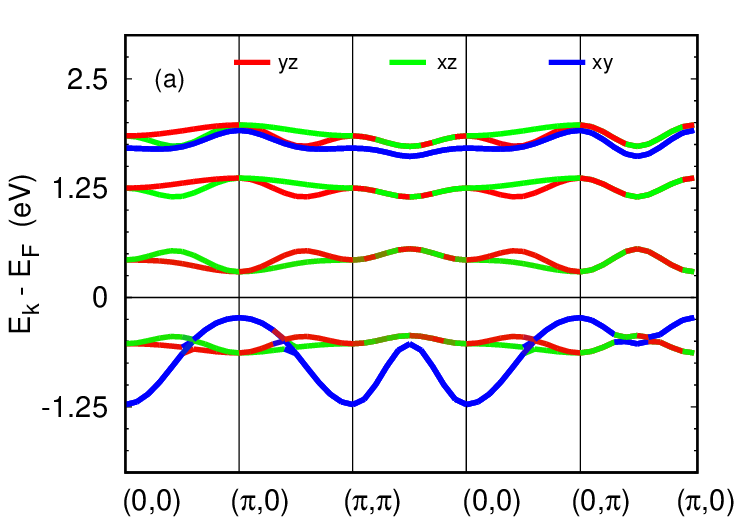,angle=0,width=75 mm,angle=0}
\psfig{figure=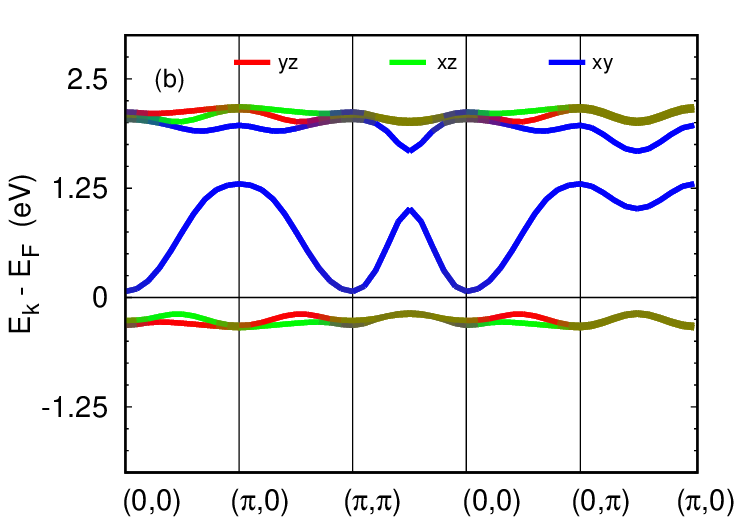,angle=0,width=75 mm,angle=0}
\caption{Orbital resolved electronic band structure calculated in the self-consistent state for: (a) $\epsilon_{xy}=-0.5$ and (b) $\epsilon_{xy}=+1.0$, with orbital and magnetic order as shown in Fig. \ref{fig2}.} 
\label{fig4}
\end{figure}

\begin{figure}[b]
\vspace*{0mm}
\hspace*{0mm}
\psfig{figure=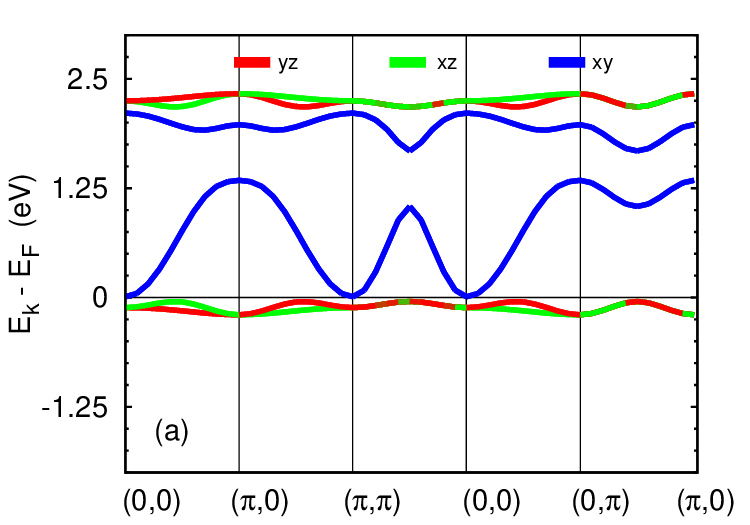,angle=0,width=75 mm,angle=0}
\psfig{figure=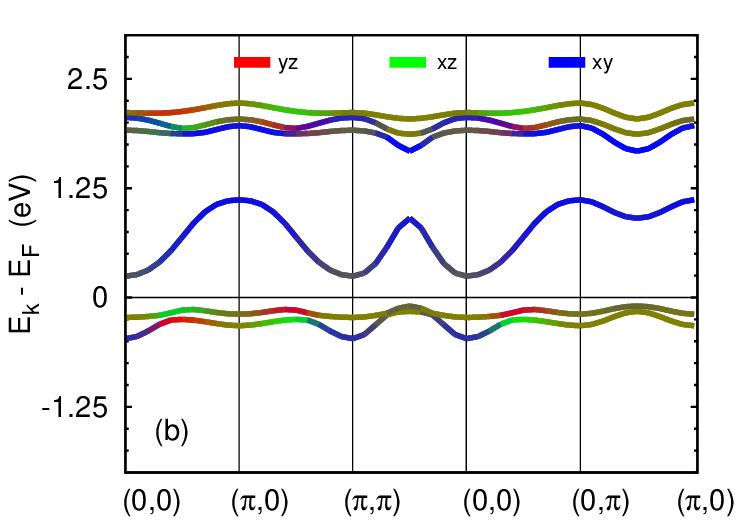,angle=0,width=75 mm,angle=0}
\caption{Orbital resolved electronic band structure in the self-consistent state: (a) without and (b) with the orbital mixing condensates included. SOC is set to zero in (a) and weak SOC has negligible effect in (b). Here $\epsilon_{xy}=0.5$.} 
\label{fig5}
\end{figure}

\subsection{Electronic band structure}
The orbital resolved electronic band structure calculated in the self-consistent state is shown in Fig. \ref{fig4} for the two orders shown in Fig. \ref{fig2}. In both cases, the large gaps across the Fermi energy $E_{\rm F}$ are due to Hubbard $U$ between opposite-spin bands for the three orbitals. The small gaps between same-spin bands are due to: (a) staggered orbital ($yz/xz$) ordering and (b) crystal field term $\epsilon_{xy}$. The number of $yz/xz$ bands in (a) is doubled compared to (b) due to staggered orbital ordering. 

\subsubsection*{Role of orbital mixing terms on electronic band structure}
We now highlight the important role of Coulomb interaction induced orbital mixing terms (Eqs. \ref{h_hf_od},\ref{sc_od}). When OM condensates and SOC are neglected (standard HF approach), bands for the three orbitals are completely decoupled due to absence of orbital mixing. For $\epsilon_{xy}=0.5$, the $xy$ band lies just above the $yz,xz$ bands as seen in Fig. \ref{fig5}(a). However, when the OM condensates are included (generalized HF approach), the band structure [Fig. \ref{fig5}(b)] shows a significant insulating gap induced by the Coulomb orbital mixing terms. Also, part of the $xy$ band is now seen to lie below the $yz,xz$ bands. With $\epsilon_{xy}$ decreasing further, $xy$ orbital weight is progressively transferred below the insulating gap, until the entire $xy$ band is shifted below the Fermi energy [Fig. \ref{fig4}(a)], while the insulating gap remains finite. 


It is important to note that the orbital mixing induced insulating gap and the electronic band structure [Fig. \ref{fig5}(b)] are nearly unchanged in the weak SOC regime. This is because the OM condensates are generated spontaneously from the Coulomb interaction terms. In the absence of SOC, these condensates are purely real, and acquire imaginary part (rotation in the complex plane) when SOC is introduced, resulting in finite orbital moments and SOC renormalization (Eq. \ref{phys_quan}). Here, the role of SOC is crucial, because finite circulating orbital and spin-orbital currents are generated due to spin-orbital entanglement, resulting in finite orbital and spin-orbital moments.  

\subsection{Collective excitations}
\begin{figure}
\vspace*{0mm}
\hspace*{0mm}
\psfig{figure=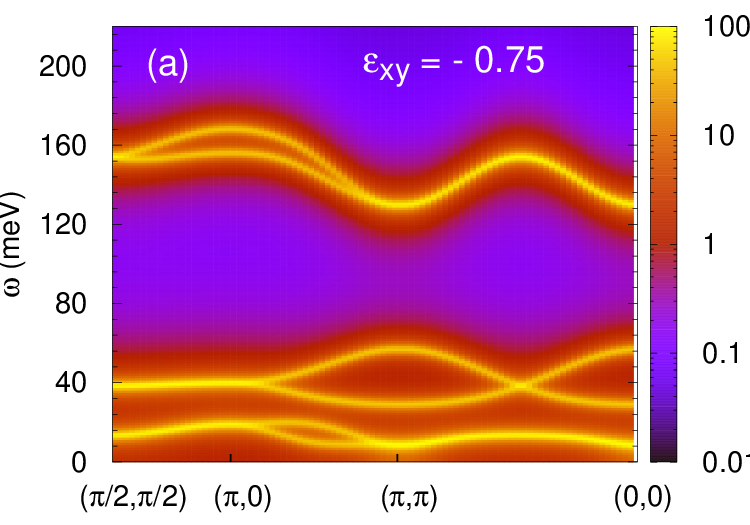,angle=0,width=53mm,angle=0}
\psfig{figure=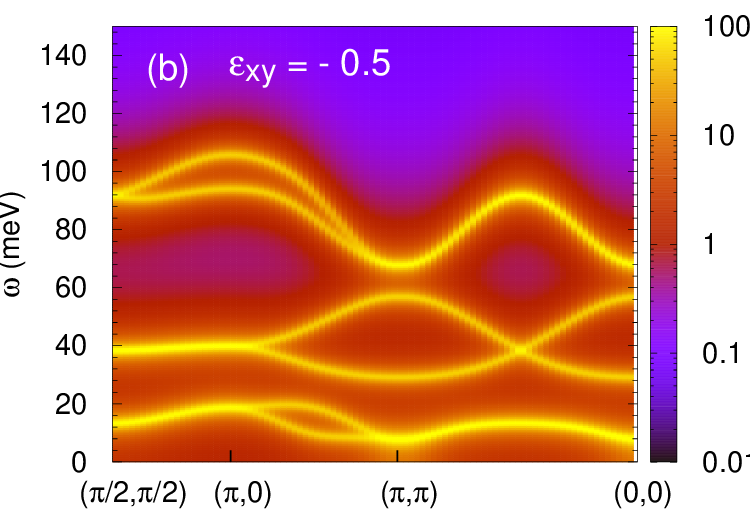,angle=0,width=53mm,angle=0}
\psfig{figure=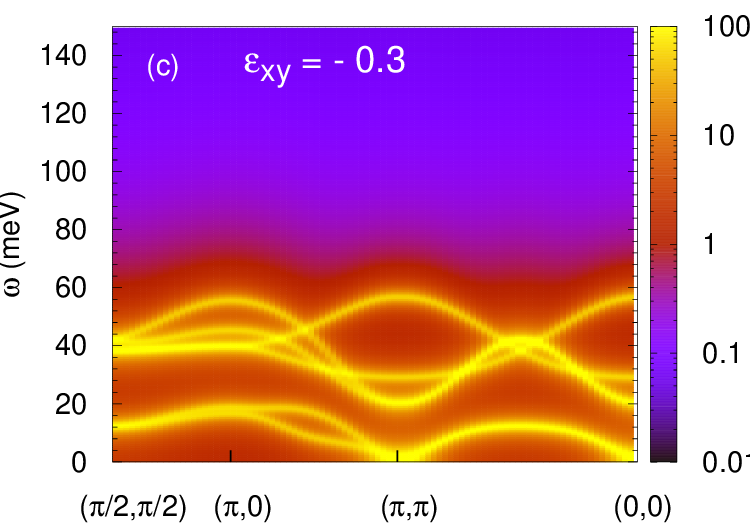,angle=0,width=53mm,angle=0}
\caption{Low-energy part of the spectral function in the axial ordering case for several $\epsilon_{xy}$ values, showing magnon (below 20 meV) and two orbiton modes. While energy of the lower orbiton mode (involving $yz,xz$ orbitals) remains unchanged, energy of the upper orbiton mode (involving $xy$ and $yz/xz$ orbitals) decreases with decreasing crystal field splitting. Here $J_{\rm H}=U/5.5$.} 
\label{fig6}
\end{figure}

In the axial ordering case (negative $\epsilon_{xy}$, staggered orbital order), low-energy part of the spectral function calculated for several $\epsilon_{xy}$ values shows (Fig. \ref{fig6}) magnon modes (below 20 meV) and two orbiton modes. As expected, energy of the lower orbiton mode (not involving $xy$ orbital) remains constant, while that of the upper orbiton mode (involving $xy$ hole) decreases from 160 meV to 40 meV with decreasing $|\epsilon_{xy}|$. Slightly higher $J_{\rm H}$ value was taken here in order to separate the lower orbiton and magnon modes for clarity, which are otherwise nearly overlapping in energy for $J_{\rm H}=U/6$. 


In addition, there are three spin-orbiton modes ($\sim$ 400, 500, 625 meV for $\epsilon_{xy}=-0.5$) and high-energy magnon mode ($\sim$ 650 meV) which reflects the cost of out-of-phase spin fluctuations of different orbitals due to Hund's coupling. The small magnon gap (8 meV) reflects the SOC induced easy ($z$) axis anisotropy and finite energy cost for transverse spin fluctuations. The low magnon energy is due to frustration and competing (FM) order preferred by $yz,xz$ moments. Consequently, magnon energy increases with decreasing $J_{\rm H}$, and the orbiton mode energy decreases due to enhanced interaction term $U''-J_{\rm H}/2$, which lowers the orbiton mode energy in the usual resonant scattering mechanism.

\vspace*{0mm}
\begin{figure}[b]
\hspace*{0mm}
\psfig{figure=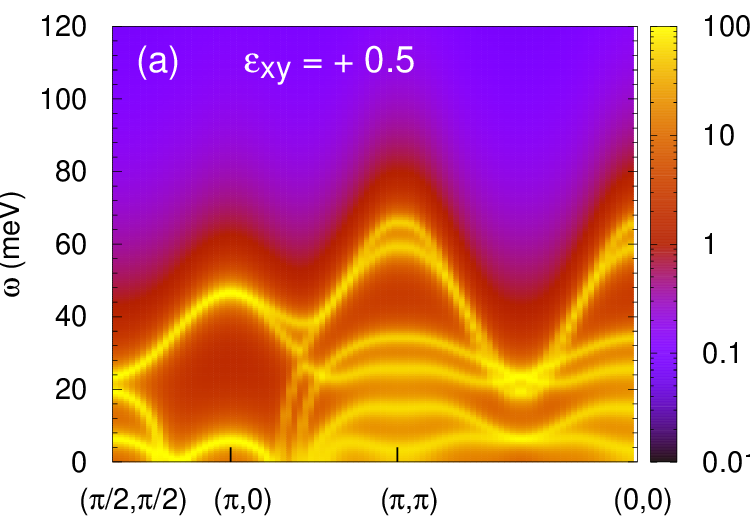,angle=0,width=50mm}
\psfig{figure=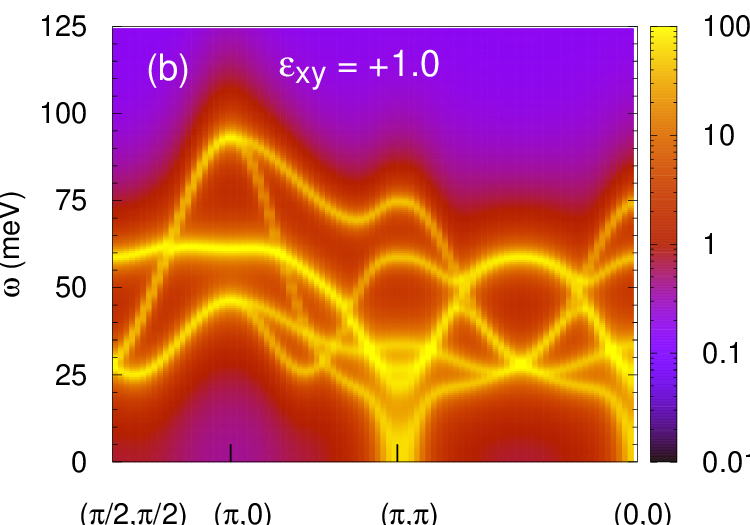,angle=0,width=50mm}
\psfig{figure=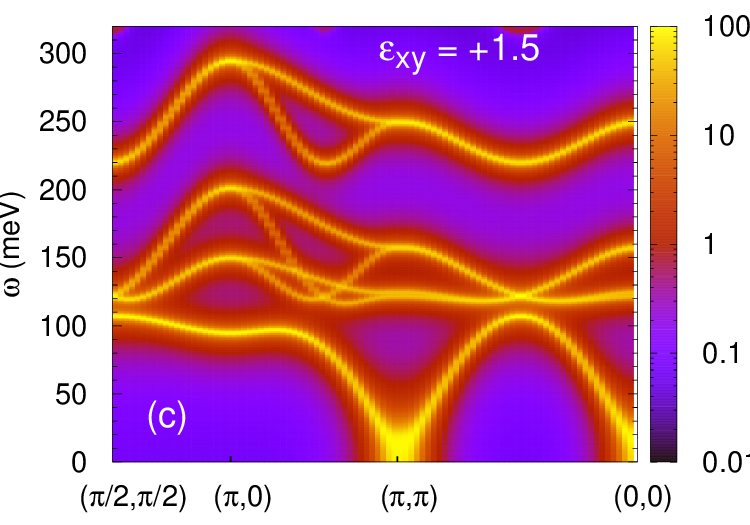,angle=0,width=50mm}
\caption{Spectral function in the planar ordering case, showing magnon and orbiton modes for positive $\epsilon_{xy}$. While energy of the orbiton mode (involving $xy$ orbital) increases with crystal field, the magnon mode energy increases only slightly. (b) shows magnon-orbiton coupling when energy of the two modes are comparable.} 
\label{fig7}
\end{figure}

In the planar ordering case (positive $\epsilon_{xy}$), the calculated spectral function is shown in Fig. \ref{fig7}. Again, the most significant feature is that while energy of the orbiton mode  increases with $\epsilon_{xy}$ due to increasing particle-hole excitation energy, the magnon mode energy increases only slightly. The orbiton modes arise from particle-hole excitations involving $xy$ (particle) and $yz,xz$ (hole) orbitals, and therefore correspondng to $L_x,L_y$ fluctuations. The prominent gapped magnon mode (b) shows a gap of 25 meV at ${\bf q}=(\pi,\pi)$ for out-of-plane spin fluctuations. The gap for the in-plane magnon mode (b) is about 3 meV. When the $xy$ band is pushed up by increasing $\epsilon_{xy}$, the orbiton modes shift to higher energy (c) and decouple from the magnon modes, which are now reduced to conventional magnons in two-dimensional quantum Heisenberg AFM model with effective spin $S=1$ constituted by Hund's coupled $yz,xz$ moments.


\section{Entangled Orbital Order ($\lambda > \lambda^*$)}
In the reversed crystal field regime $(\epsilon_{xy}\sim -1)$, where orbital degree of freedom is maximal in the $yz/xz$ sector since $n_{yz} + n_{xz} \sim 1$, we obtain a transition from staggered orbital order (SOO) to entangled orbital order (EOO) at a critical SOC value ($\lambda^*$) which is $U$ and $J_{\rm H}$ dependent. In the following, we will consider the realistic parameter set: $U=12$ (3 eV), $J_{\rm H} = U/6$, and $\epsilon_{xy}=-0.75$ unless otherwise indicated, with the hopping parameters and energy scale ($|t_1| = 250$ meV) same as earlier. For the bare SOC, we will consider values upto $\lambda=0.2$, which is above the critical value obtained $\lambda^*=0.13$ (30 meV). 

\begin{figure}
\vspace*{0mm}
\hspace*{0mm}
\psfig{figure=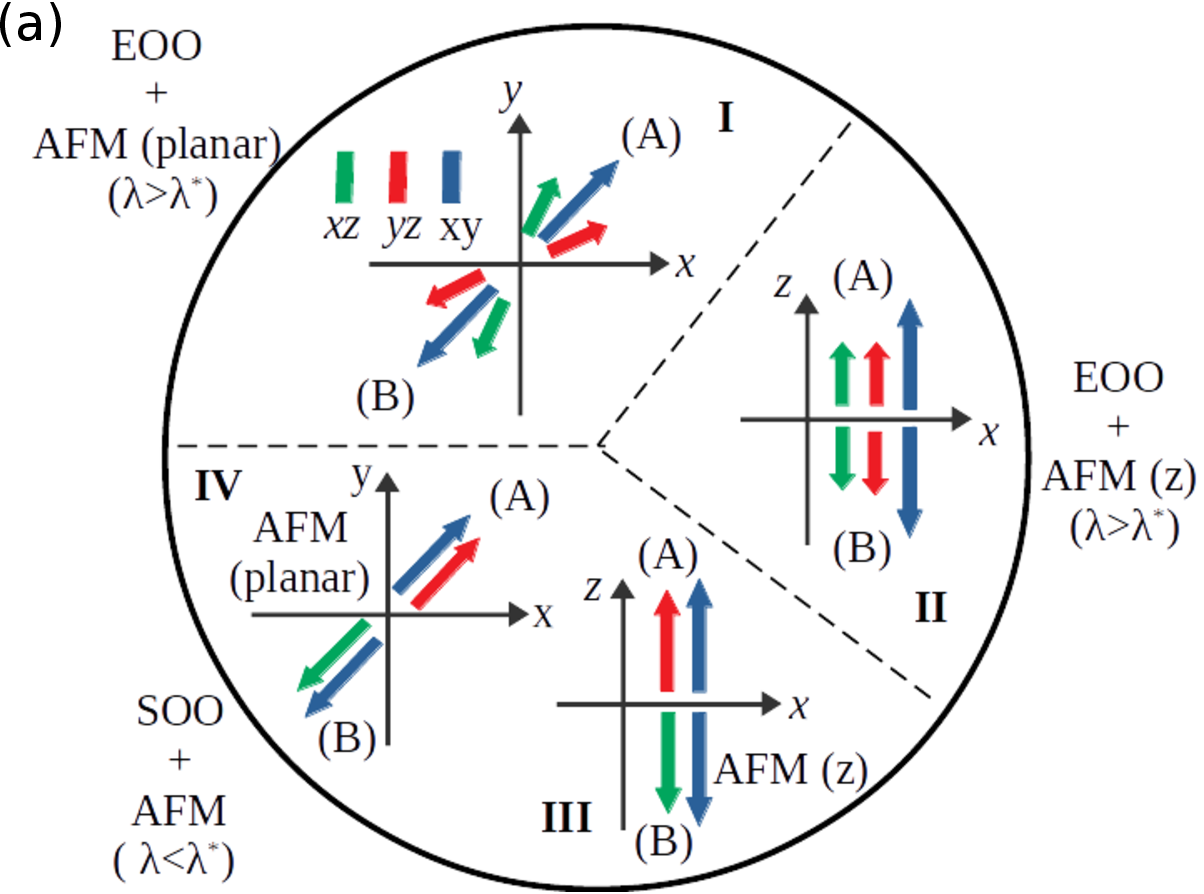,angle=0,width=70 mm,angle=-0}
\psfig{figure=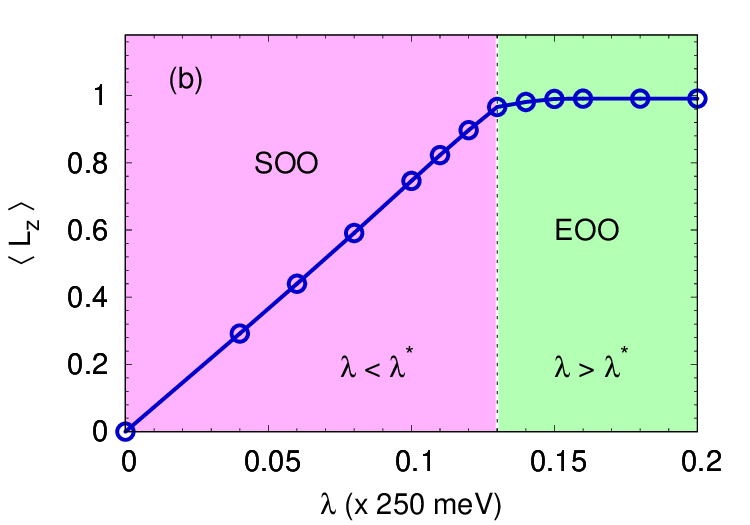,angle=0,width=74 mm,angle=0} \\ \vspace*{3mm}
\psfig{figure=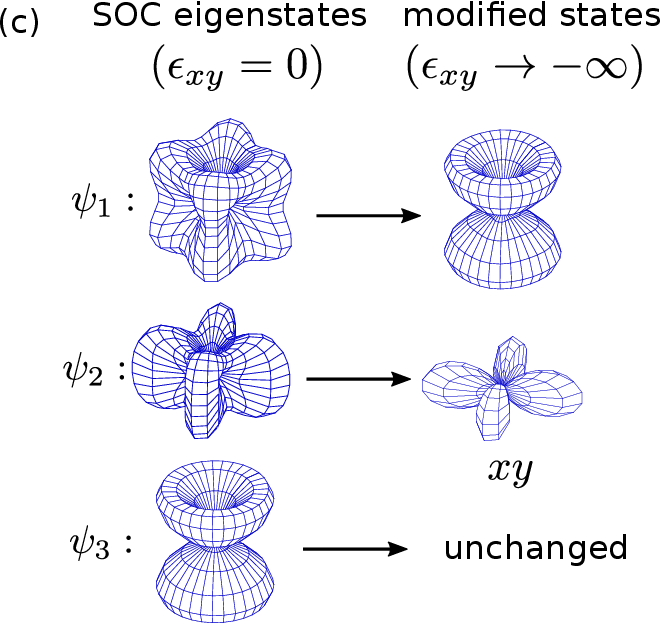,angle=0,width=50 mm,angle=0} \hspace*{10mm}
\psfig{figure=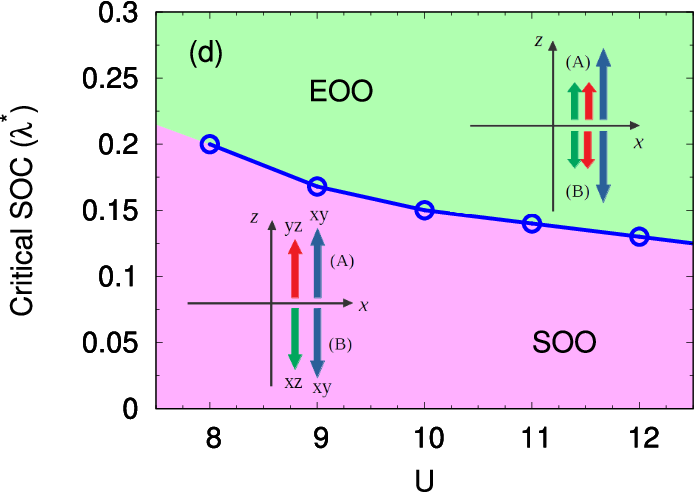,angle=0,width=70 mm,angle=-0}
\caption{(a) Schematic diagram showing self-consistent orders obtained. Arrow size, orientation, color indicate spin moment magnitude, orientation, orbital. The transitions obtained are: SOO to EOO (IV to I and III to II) when SOC is increased beyond critical value $\lambda^*$, reorientation due to SOC induced anisotropy (IV to III and I to II) when a small perturbation $\delta m_z$ (AFM) is included, and reorientation + SOO to EOO (IV to II) when both $\delta m_z$ and increased SOC are included. (b) The saturation of orbital moment $\langle L_z \rangle = 1$ for $\lambda > \lambda^*$ reflects the optimal entanglement of $yz,xz$ orbitals. (c) Orbital shapes of the SOC eigenstates. (d) Variation of $\lambda^*$ with $U$ showing the phase boundary between SOO and EOO. Here $\epsilon_{xy} = -0.75$ and $J_{\rm H} = U/6$.} 
\label{fig8}
\end{figure}

Results of the self-consistent calculation are summarized in Fig. \ref{fig8}(a), schematically showing the magnitude and orientation of spin moments for different orbitals obtained in the entangled orbital order ($n_{yz}=n_{xz}$) for $\lambda > \lambda^*$, and in the staggered orbital order ($n_{yz} \ne n_{xz}$) for $\lambda < \lambda^*$ as discussed in the previous section. Final self consistent magnetic order obtained is AFM ($z$) in both cases. The optimal entanglement in the $yz/xz$ sector is reflected by the saturated orbital moment $\langle L_z \rangle = 1$ for $\lambda > \lambda^*$, as shown in Fig. \ref{fig8}(b).

To understand the nature of the EOO, it is convenient to consider the entangled eigenstates of the SOC + crystal field Hamiltonian:
\begin{eqnarray}
\label{socstate1}
\psi_1 &=& \frac{1}{\sqrt{2+\alpha^2}} \left[ \sigma_x \psi_{yz} +\sigma_y \psi_{xz} + \alpha \sigma_z \psi_{xy} \right] \nonumber \\
\psi_2 &=& \frac{1}{\sqrt{2+\beta^2}} \left[\sigma_x \psi_{yz} +\sigma_y \psi_{xz} - \beta \sigma_z \psi_{xy} \right] \\
\psi_3 &=& \frac{1}{\sqrt{2}} \left [\sigma_x \psi_{yz} -\sigma_y \psi_{xz} \right]  \nonumber
\end{eqnarray}
where the coefficients are given by:
\begin{eqnarray}
\alpha &=& \frac{-1 + \zeta + \sqrt{9+\zeta^2 -2\zeta}}{2} \nonumber \\
\beta &=& \frac{1 - \zeta + \sqrt{9+\zeta^2 -2\zeta}}{2}
\end{eqnarray}
in terms of the parameter $\zeta = 2\epsilon_{xy} /\lambda$. For negative $\epsilon_{xy}$, in the limit $\zeta \rightarrow -\infty$, where $\alpha \approx 0$ and $\beta \approx |\zeta|$, the eigenstates reduce to:
\begin{eqnarray}
\psi_1 &\approx& \frac{1}{\sqrt{2}} \left[ \sigma_x \psi_{yz} +\sigma_y \psi_{xz} \right] \nonumber \\
\psi_2 &\approx&  -\sigma_z \psi_{xy} \nonumber \\
\psi_3 &=& \frac{1}{\sqrt{2}} \left [\sigma_x \psi_{yz} -\sigma_y \psi_{xz} \right] 
\label{socstate2}
\end{eqnarray}
which are shown in Fig. \ref{fig8}(c). The two eigenstates $\psi_2$ and $\psi_3$ belonging to the total angular momentum $J=3/2$ sector have the lowest energy. The AFM order corresponds to occupation of $\psi_2$ and $\psi_3$ states with pseudo-spin $\tau=\uparrow$ on A and $\tau=\downarrow$ on B sublattices. The opposite pseudo-spin cases for $\psi_2$ and $\psi_3$ along with the $\psi_1$ pseudo-spin doublet are the four unoccupied states above the Fermi energy. This picture is confirmed by the electronic band structure (Fig. \ref{fig10}) which shows the six emergent bands. 



\begin{figure}
\vspace*{0mm}
\hspace*{0mm}
\psfig{figure=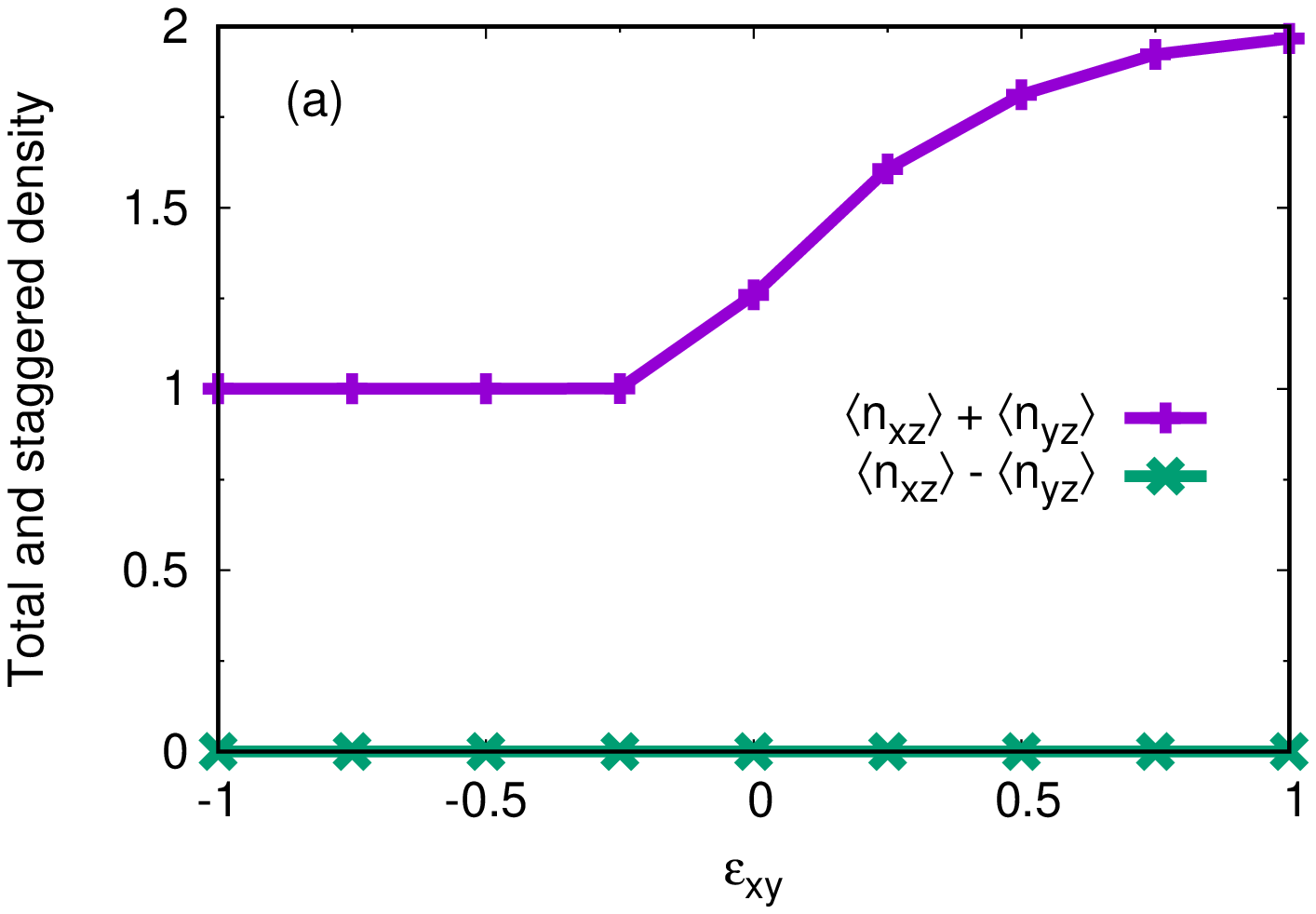,angle=0,width=40 mm,angle=-90}
\psfig{figure=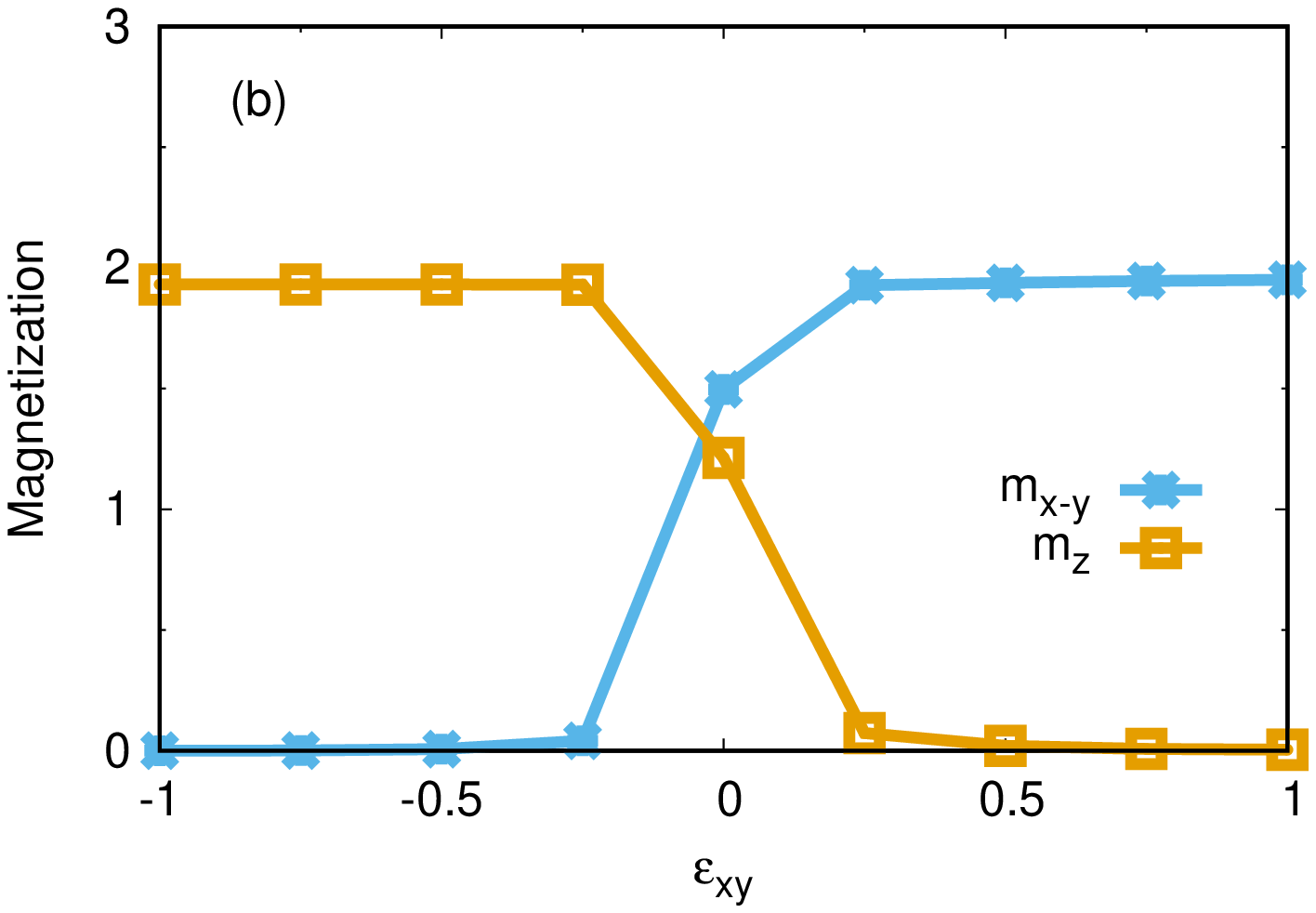,angle=0,width=40 mm,angle=-90}
\psfig{figure=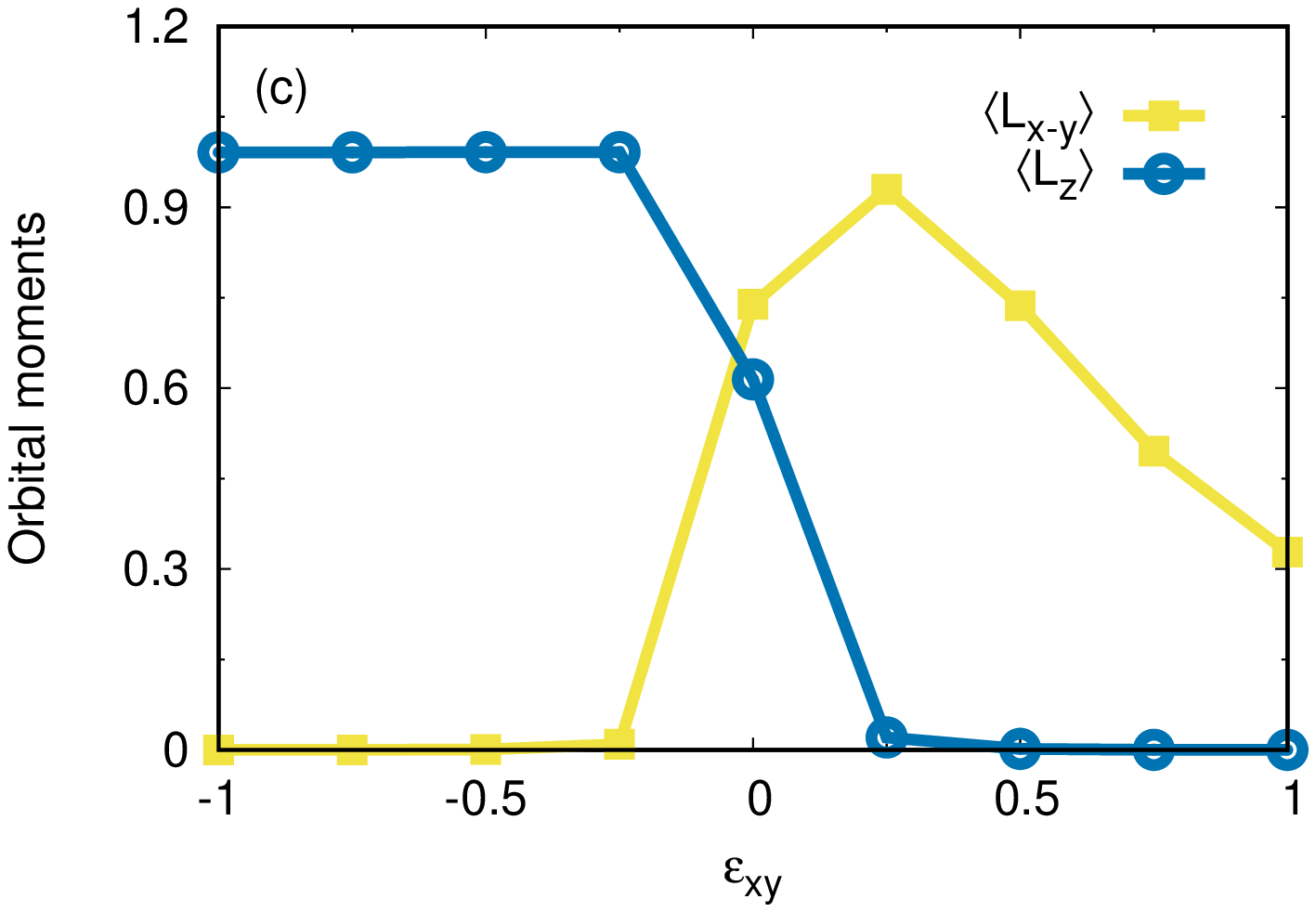,angle=0,width=40 mm,angle=-90}
\psfig{figure=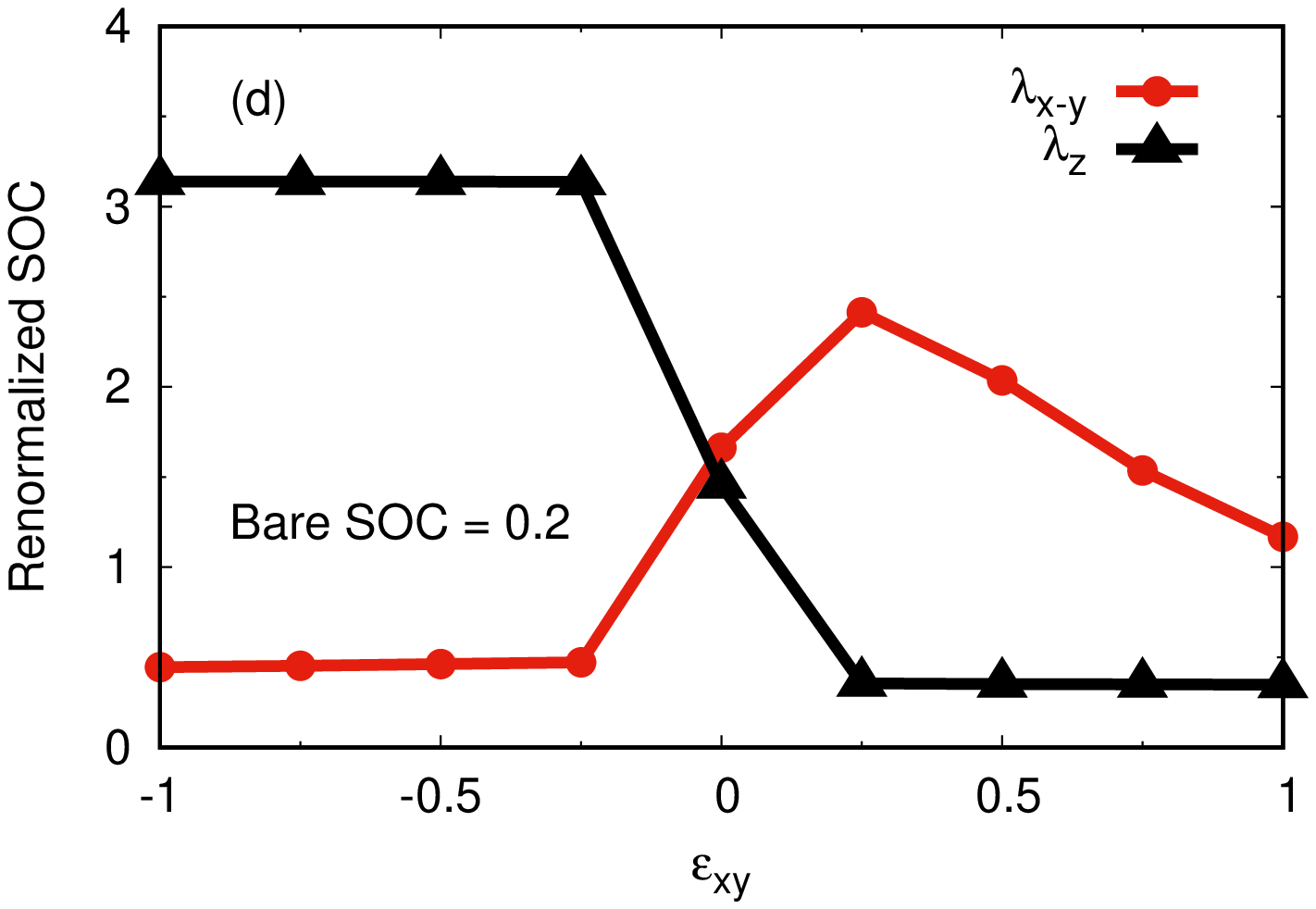,angle=0,width=40 mm,angle=-90}
\caption{Evolution of various physical quantities with crystal field. (a) The total and differential electron occupancies in the $yz/xz$ sector show the entangled orbital order ($n_{yz}-n_{xz}=0$) even when $n_{yz} + n_{xz}=1$. (b) Axial and planar magnetic moments and (c) orbital moments, showing the magnetic reorientation transition. The orbital moments (c) and Coulomb renormalized SOC values (d) are stronger in comparison to the SOO case.} 
\label{fig9}
\end{figure}

Fig. \ref{fig8}(d) shows the variation of the critical SOC value with $U$ for fixed Hund's coupling ($J_{\rm H} = U/6$). The critical SOC value weakly decreases with increasing $U$ (which suppresses band effects) and with decreasing $J_{\rm H}$ (not shown), because both band effects (hopping terms mix different $J$ sectors) and $J_{\rm H}$ (which tends to align moments of different orbitals) are detrimental to the SOC induced entanglement. The critical SOC value remains unaffected by $\epsilon_{xy}$ in the considered regime.

For SOC value $\lambda=0.2$ (EOO regime), results of the generalized self consistent calculation showing the evolution of various physical quantities of interest with the crystal field term in the range $-1 \le \epsilon_{xy} \le +1$ are presented in Fig. \ref{fig9}. The magnetic and orbital moments are seen to smoothly reorient from the axial to planar direction due to SOC induced magnetic anisotropy. As expected, the orbital moments and renormalized SOC values are stronger compared to the SOO case (Fig. \ref{fig1}) due to enhanced SOC induced orbital entanglement. 

The calculated electronic band structure and magnetic excitation spectral function in the AFM state with EOO are shown in Fig. \ref{fig10}. The band structure is much simpler in this case with only six doubly degenerate bands and no fine splitting between $yz/xz$ bands as in the SOO case (Fig. \ref{fig4}(a)). The three doubly degenerate bands formed near the Fermi energy are derived from the three entangled states discussed above. Similarly, three bands for opposite spin are formed at much higher energy due to the Hubbard interaction $U$. The insulating gap in this case is produced by the Coulomb orbital mixing terms, while it was the staggered orbital field which was responsible in the SOO case.

\begin{figure}
\hspace*{0mm}
\psfig{figure=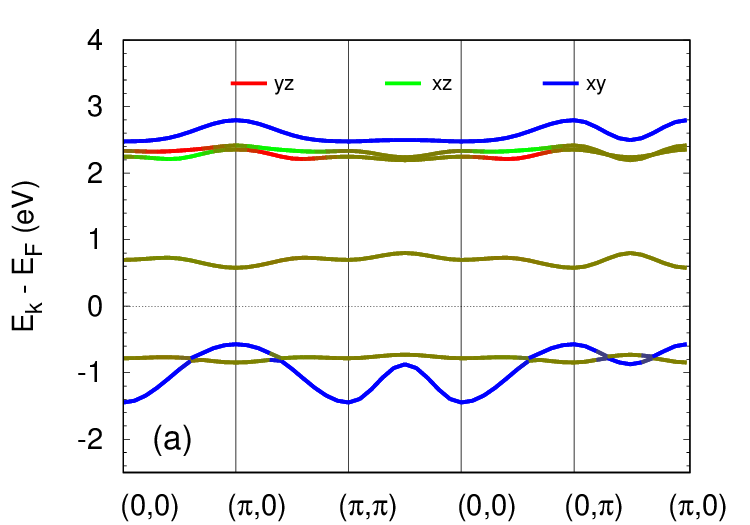,angle=0,width=65 mm,angle=0}\hspace{10mm}
\psfig{figure=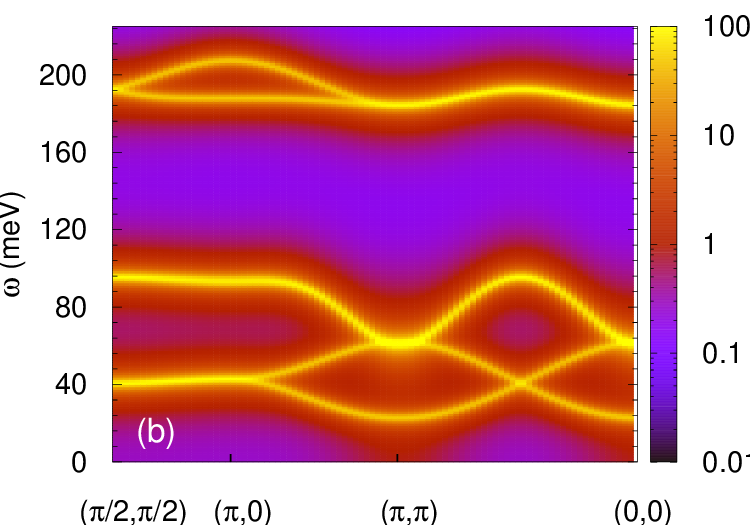,angle=0,width=65 mm,angle=0}
\caption{(a) The orbital resolved electronic band structure and (b) low energy part of the spectral function, both calculated in the AFM + EOO for bare SOC $\lambda=0.2$.} 
\label{fig10}
\end{figure}

The lowest energy collective excitation ($\sim 40$ meV) in Fig. \ref{fig10}(b) corresponds to the $yz/xz$ orbiton mode involving particle-hole excitation between bands originating from the entangled $\psi_1$ and $\psi_3$ states discussed above. The second mode ($\sim 90$ meV) is the magnon mode corresponding to transverse spin fluctuations away from the easy $z$ axis. The large magnon gap ($\sim 60$ meV) reflects strong magnetic anisotropy resulting from highly enhanced Coulomb renormalized SOC as seen in Fig. \ref{fig9}(d). The third pair ($\sim 200$ meV) are orbiton modes corresponding to p-h excitations involving $yz/xz$ (particle) and $xy$ (hole) bands originating from the $\psi_1$ and $\psi_2$ states.

\begin{figure}
\vspace*{0mm}
\hspace*{0mm}
\psfig{figure=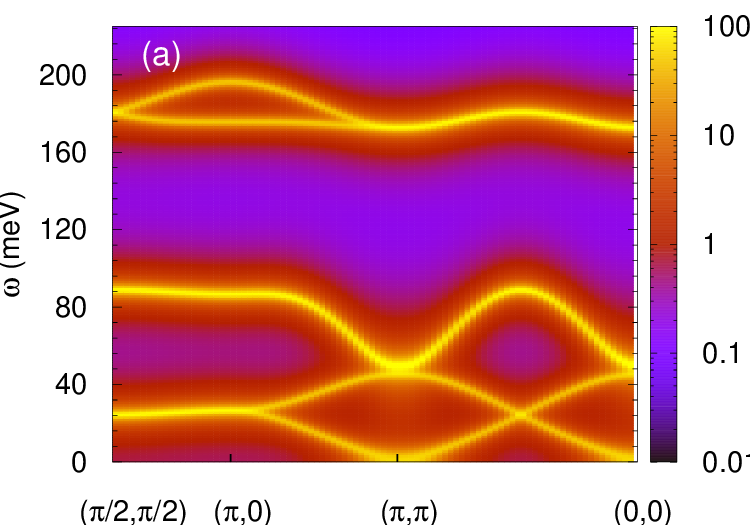,angle=0,width=65 mm,angle=0}\hspace{10mm}
\psfig{figure=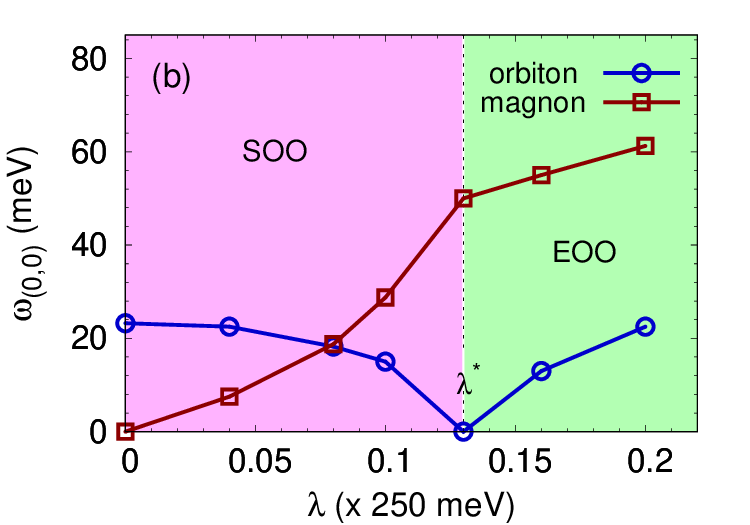,angle=0,width=65 mm,angle=0}
\caption{(a) Spectral function in the EOO + AFM order shows that the lower orbiton mode becomes gapless at the critical SOC value $\lambda^* = 0.13$, reflecting degeneracy with SOO case. (b) Variation of both orbiton and magnon gap energies at ${\bf q}=0$ with bare SOC strength.} 
\label{fig11}
\end{figure}

When SOC strength reduces to the critical value, the $yz/xz$ orbiton mode becomes gapless [Fig. \ref{fig11}(a)] due to the degeneracy between staggered and entangled orbital orders. Fluctuations which change the orbital order from entangled to staggered or vice versa cost no energy for ${\bf q}=0$. The orbiton gap becomes finite on both sides of $\lambda^*$, as seen in Fig. \ref{fig11}(b). Also shown is the variation of the calculated magnon gap, which properly goes to zero with SOC as expected.

\begin{figure}[b]
\vspace*{0mm}
\hspace*{0mm}
\psfig{figure=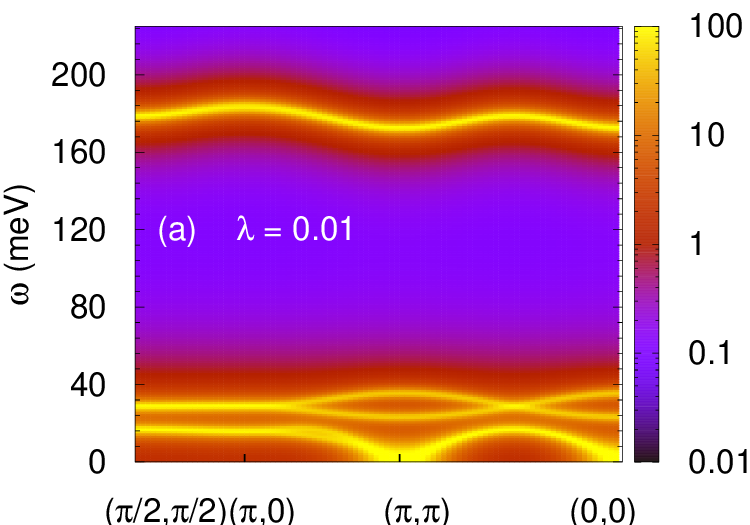,angle=0,width=53 mm,angle=0}
\psfig{figure=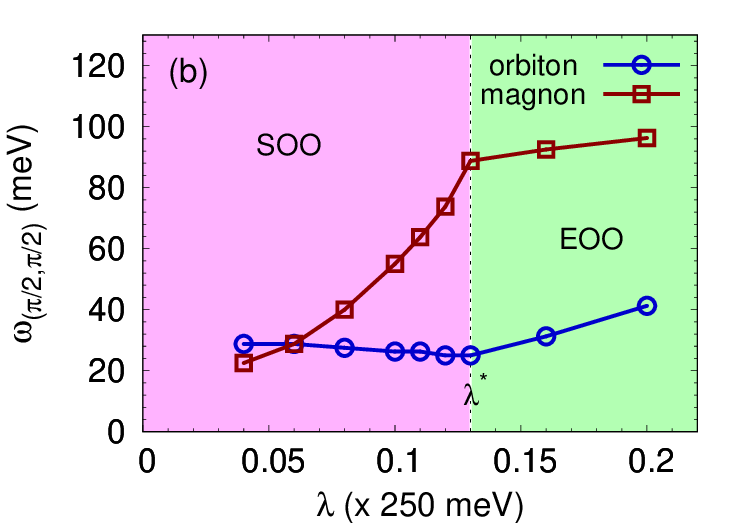,angle=0,width=53 mm,angle=0}
\psfig{figure=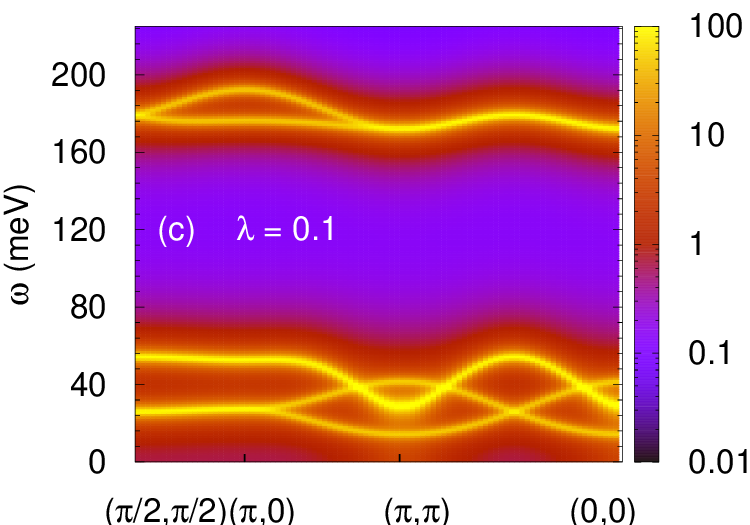,angle=0,width=53 mm,angle=0}
\caption{The spectral function of collective excitations calculated for the SOO case with SOC values: (a) $\lambda=0.01$ and (c) $\lambda=0.1$. (b) Variation of magnon and orbiton mode energy scales with bare SOC strength. Here $\epsilon_{xy}=-0.75$.} 
\label{fig12}
\end{figure}

The strong contrast between the weak SOC ($\lambda \sim 0.1$) and no SOC cases is highlighted in Fig. \ref{fig12} which shows the spectral functions in the two cases. The energy of the magnon mode is lowest in (a) but increases significantly above the orbiton mode in (c). The variation of magnon and orbiton mode energy scales ($\omega_{\bf q}$ for ${\bf q}=(\pi/2,\pi/2)$) with SOC strength is shown in (b). Even in the SOO case (below critical SOC strength), there is strong enhancement of the magnon energy scale with SOC. This shows that the weak SOC case is remarkably different from the no SOC case. 

\begin{figure}
\vspace*{0mm}
\hspace*{0mm}
\psfig{figure=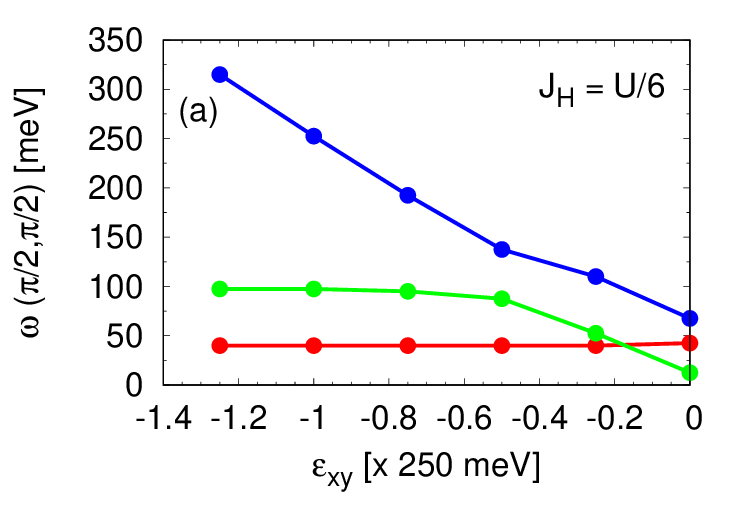,angle=0,width=53 mm,angle=0}
\psfig{figure=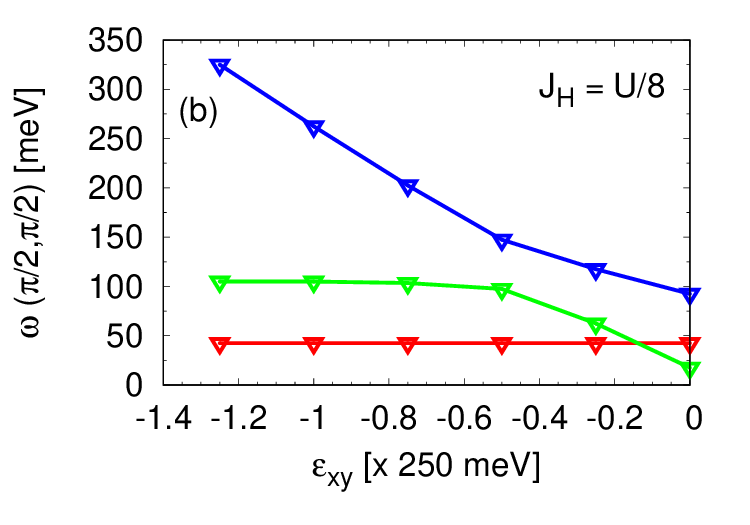,angle=0,width=53 mm,angle=0}
\psfig{figure=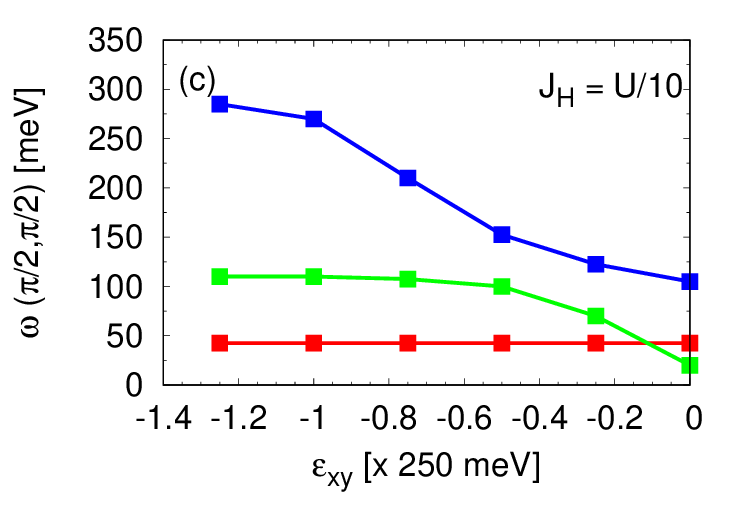,angle=0,width=53 mm,angle=0} \vspace{-5mm}
\caption{Variation of the different collective excitation energies at ${\bf q}=(\pi/2, \pi/2)$ with crystal field in the EOO + AFM order for several $J_{\rm H}$ values.} 
\label{fig13}
\end{figure}

Hund's coupling only weakly affects the collective excitation energy scales. Fig. \ref{fig13} shows the variation of the calculated excitation energies at ${\bf q}=(\pi/2, \pi/2)$ with $\epsilon_{xy}$ in the EOO + AFM order for three $J_{\rm H}$ values. In all cases, the lower orbiton energy remains flat and the magnon energy is flat over a broad range ($\epsilon_{xy}\le -0.5$). Energy of the upper orbiton (involving $xy$ hole) decreases approximately linearly with $|\epsilon_{xy}|$.

\begin{figure}[b]
\vspace*{0mm}
\hspace*{0mm}
\psfig{figure=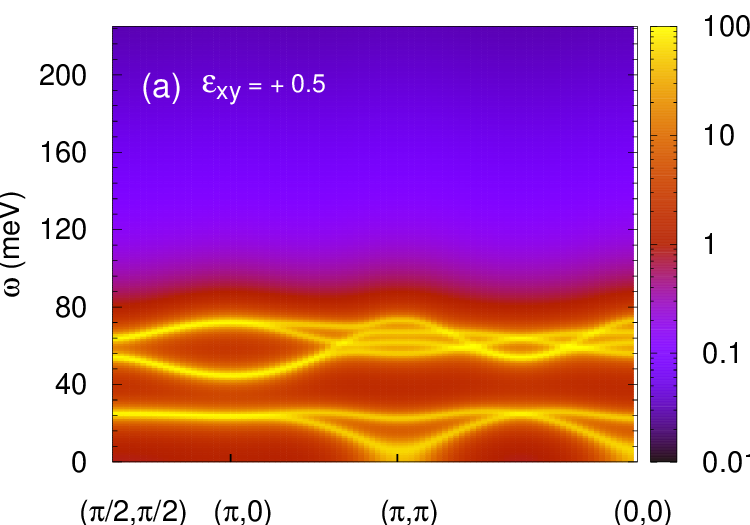,angle=0,width=53 mm,angle=0}
\psfig{figure=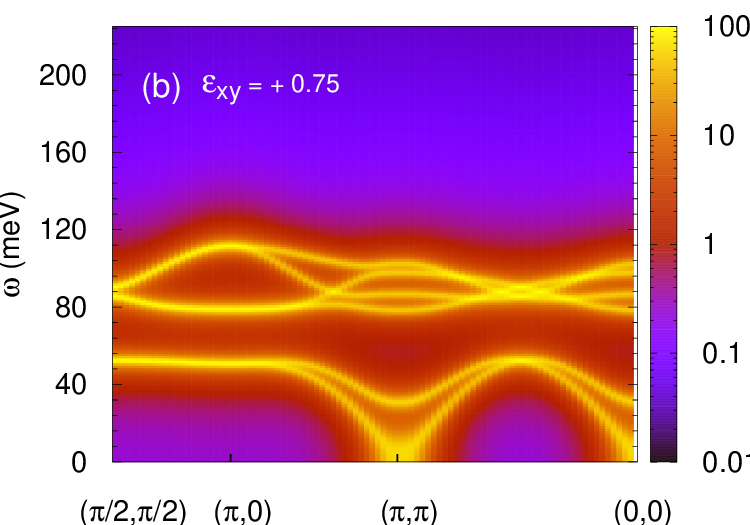,angle=0,width=53 mm,angle=0}
\psfig{figure=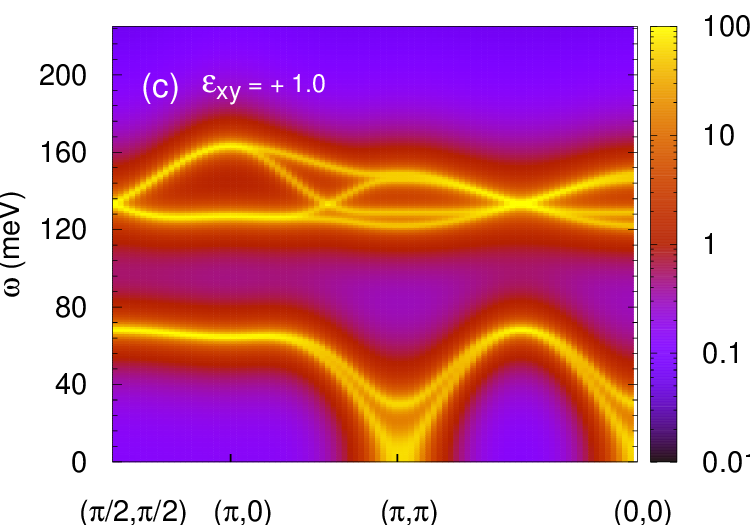,angle=0,width=53 mm,angle=0}
\caption{Spectral functions for the planar AFM order, showing magnon and orbiton modes for several positive $\epsilon_{xy}$ values: (a) 0.5, (b) 0.75, and (c) 1.0.} 
\label{fig14}
\end{figure}


Turning now to collective excitations in the planar AFM ordering case $(\epsilon_{xy}>0)$, the calculated spectral functions are shown in Fig. \ref{fig14} for several $\epsilon_{xy}$ values. The lower energy modes are the magnon modes involving the magnetically active $yz,xz$ orbitals, showing nearly gapless and strongly gapped modes corresponding to in-plane and out-of-plane spin fluctuations due to easy-plane anisotropy. The higher energy modes (energy increasing with $\epsilon_{xy}$) are the orbiton modes involving $xy$ (particle) and $yz/xz$ (hole) excitations.
 
We now discuss the estimation of N\'{e}el temperature $T_{\rm N}$ from the magnon energy scale within the picture of thermal excitation of magnons as driving the demagnetization. From Fig. \ref{fig14}(c), the magnon energy scale $\omega_{\rm max} \sim 70$ meV for $\epsilon_{xy}=+1$ corresponding to the high pressure side. Based on the well-studied finite-temperature spin dynamics including interlayer hopping and magnetic anisotropy effects,\cite{singh_PRL_1990,singh_PRB_1991} qualitatively similar magnon dispersion in the cuprate compound $\rm La_2CuO_4$ with $T_{\rm N} \sim 400$ K and $\omega_{\rm max} \sim 300$ meV,\cite{singh_PRB_2002} and neglecting differences in interlayer hopping and magnetic anisotropy gaps which only weakly (logarithmically) affect the spin dynamics, we obtain $T_{\rm N} \sim 90$ K which is close to the measured transition temperature $T_{\rm N} \sim 120$ K in the chromate compound as seen on the high-pressure side of the $P-T$ diagram.\cite{yamauchi_PRL_2019}

\section{Discussion}

Orbital moments $\langle L_\alpha \rangle$ and spin-orbital moments $\langle L_\alpha S_\beta \rangle$ (where $\alpha,\beta=x,y,z$) are generated by circulating orbital and spin-orbital currents which are finite only in presence of SOC induced spin-orbital entanglement. Weak SOC in strongly correlated $3d$ transition metal systems is usually neglected in theoretical studies, and therefore consideration of orbital and spin-orbital moments is naturally excluded. Also generally neglected are the Coulomb orbital mixing terms as in Eqs. \ref{h_hf_od},\ref{sc_od} which are generated in the generalized self consistent approach. Furthermore, in studies where staggered $yz/xz$ ordering is attributed to Jahn-Teller effect only, Coulomb renormalization of crystal field splitting is not considered due to neglect of inter-orbital density interaction term $U''$. 

The most dramatic effect of SOC seen in our calculation for the $\rm Sr_2CrO_4$ compound is the staggered to entangled orbital order transition, where the $yz/xz$ orbiton mode becomes gapless. The weak critical SOC value ($\lambda^*=0.13=30$ meV) lies in the realistic range for $3d$ elements. Other important consequences of SOC include: strong orbital and spin-orbital moments, Coulomb renormalized SOC, magnetic anisotropy, and finite magnon gap. Even for weak bare SOC, the high magnon gap energy is due to strongly enhanced Coulomb renormalized SOC. 




This work provides insight into the observed behavior of the two transition temperatures with pressure which tunes the crystal field. In the high-pressure study,\cite{yamauchi_PRL_2019} while temperature of the 140 K transition (at ambient pressure) was found to decrease linearly with pressure, that of the 110 K transition remains nearly unchanged. Both transitions (obtained from $\chi-T$ anomalies) disappear near 3 GPa where their temperatures meet. A similar but inverted behavior is obtained in the higher pressure regime, where the two transition temperatures (obtained from $\rho-T$ anomalies) approach each other with decreasing pressure and meet near 8 GPa. In addition, a peculiar insulator-to-insulator transition was obtained at around 5 GPa between the two pressure regimes. Disappearance of the 140 K transition was attributed to restoration of the conventional crystal field in a consequent DFT study.\cite{takahashi_JPSC_2020}

\begin{figure}
\vspace*{0mm}
\hspace*{0mm}
\psfig{figure=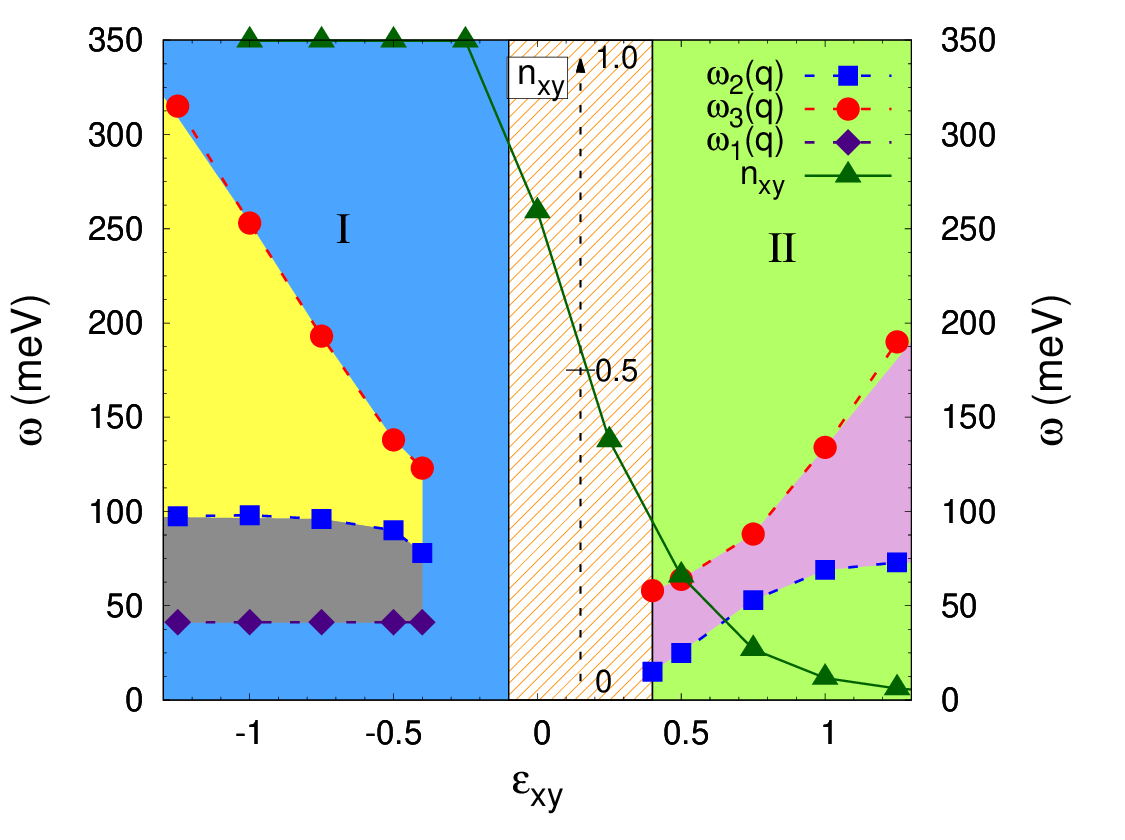,angle=0,width=90 mm,angle=0}
\caption{Variation of the excitation energy scales with $\epsilon_{xy}$ in the two regions I and II. The in-between shaded region is where the $xy$ orbital electron spectral weight transfer across the insulating gap is active. Here bare SOC value $\lambda=0.2$, $U=12$, and $J_{\rm H}=U/6$. These features closely track the observed behaviour of the two transition temperatures and the insulator-insulator transition found in the high-pressure study.\cite{yamauchi_PRL_2019}} 
\label{fig15}
\end{figure}

Towards a qualitative comparison with results of the high-pressure study, Fig. \ref{fig15} shows the behavior of relevant collective excitation energy scales: lower energy orbiton mode ($\omega_1$) involving $yz/xz$ orbitals in region I, and magnon mode ($\omega_2$) and higher energy orbiton mode ($\omega_3$) in both regions I and II. Variation of the calculated energy scales with $\epsilon_{xy}$ in the two regions is in striking similarity to the behavior of the two transition temperatures with pressure in the $P \lesssim 3$ GPa and $P \gtrsim 6$ GPa regimes, respectively. This picture is consistent with thermal excitation of collective modes driving the phase transitions. Also shown is variation of the $xy$ orbital electron density $n_{xy}$. The shaded $\epsilon_{xy}$ region, where active spectral weight transfer causes significant change in the orbital character of states across the insulating gap (see Fig. \ref{fig4}), corresponds to the insulator-insulator transition found in the high pressure study.  

In the previous section, the N\'{e}el temperature was estimated from the magnon energy scale in the positive $\epsilon_{xy}$ regime. We now consider finite-temperature effects on the EOO in the negative $\epsilon_{xy}$ regime, where the lowest-energy mode is the $yz/xz$ orbiton mode (Fig. \ref{fig15}) involving excitations between the entangled $|yz\rangle \pm i|xz\rangle$ states across the Fermi energy. Resulting from thermal excitation of this orbiton mode, the lowest temperature effects will therefore be associated with orbital dynamics, leading to reduced orbital $\langle L_z \rangle$ and spin-orbital $\langle L_z S_z \rangle$ moments, and eventually loss of entangled orbital order and disentanglement transition. 

Without orbital and spin-orbital correlations, the magnon mode energy will be significantly lowered as in Fig. \ref{fig13} for no spin-orbit coupling, resulting in strong thermal demagnetisation and concomitant magnetic disordering (N\'{e}el) transition. The orbital disentanglement and magnetic disordering transitions approximately coincide in this proposed picture, and the measured 110 K transition temperature is consistent with the $\lesssim 10$ meV orbiton gap energy for $\lambda \gtrsim \lambda^*$ (Fig. \ref{fig11}). Qualitatively similar picture of orbital disordering driven enhancement of spin fluctuations has been proposed in earlier studies of simplified models for coupled spin-orbital systems.\cite{chen_PRB_2009}

\section{Conclusions}
Due to active $yz/xz$ orbital degree of freedom in the reversed crystal field regime, even weak SOC is found to have important consequences. SOC induced magnetic anisotropy results in AFM ($z$) order at ambient pressure and AFM (planar) order at high pressure. Staggered to entangled orbital order transition is obtained at critical SOC value $\lambda^* \sim 30$ meV. Even for weak bare SOC, the high magnon gap energy is due to strongly enhanced Coulomb renormalised SOC. Due to the low orbiton gap energy near critical SOC value, finite temperature orbiton dynamics and resulting loss of orbital correlations and disentanglement transition is proposed as the driving mechanism for the magnetic (N\'{e}el) transition. 



\begin{acknowledgments}
DKS was supported through start-up research grant SRG/2020/002144 funded by DST-SERB.
\end{acknowledgments}

\end{document}